\documentclass[11pt]{CGC2}

\usepackage[english]{babel}
\usepackage{verbatim}
\usepackage{psfrag}
\usepackage{graphicx}
\usepackage{epsfig}
\usepackage{xypic}
\usepackage{tikz}

\newcommand{\Sym}{\mathrm{Sym}}
\newcommand{\Alt}{\mathrm{Alt}}

\newcommand{\GL}{\mathrm{GL}}

\newcommand{\N}{\text{$\mathbf{N}$}}
\newcommand{\MM}{\mathcal{M}}

\begin{document}

\Logo{Preprint 2010\ \ $\qquad$ CGC latex}

\begin{frontmatter}

\title{A possible intrinsic weakness of AES and other cryptosystems}

{\author{Anna Rimoldi}} {{\tt (rimoldi@science.unitn.it)}}\\
{{Department of Mathematics, University of Trento, Italy.}}

{\author{Massimiliano Sala}} {{\tt (maxsalacodes@gmail.com)}}\\
{{Department of Mathematics, University of Trento, Italy.}}

{\author{Ilia Toli}} {{\tt (ilia.toli@gmail.com)}}\\
{{Department of Mathematics, Northeastern University, Boston, USA.}}

\runauthor{A.~Rimoldi, M.~Sala, I.~Toli}

\begin{abstract}
It has been suggested that the algebraic structure of AES (and other similar block ciphers) 
could lead to a weakness exploitable in new attacks. In this paper, we use the
algebraic structure of AES-like ciphers to construct a novel cipher embedding where the ciphers may lose 
their non-linearity. We show some examples and we discuss the limitations of our approach.
\end{abstract}

\end{frontmatter}
\section*{Introduction}
The Advanced Encryption Standard (AES) \cite{CGC-misc-nistAESfips197} is nowadays the most widespread block cipher in commercial applications. 
It represents the state-of-art in block cipher design and provides an unparalleled level of assurance against all known cryptanalytic techniques, 
except for its round-reduced versions.
It is true that AES  (and other modern block ciphers) presents a highly algebraic structure, leading researchers to exploit it for new algebraic
attacks, but these tries have been unsuccessful as yet (except for academic reduced versions).

The best that one can hope for a cryptosystem is that all its encryption functions behave in unpredictable way (close to random), 
in particular we would like that it behaves in a way totally different from linear or affine maps.

A sign of strength for AES is that nobody has been able to show that its encryption functions are any closer to linear maps than arbitrary 
random functions.

However, it might be possible to extend AES to act on bigger spaces, in such a way that the non-random behavior of AES becomes easier to spot.
For example, it was hoped that embedding AES into BES would allow easier\footnote{{\it easier} than systems coming from random maps.} 
polynomial systems to break the ciphers (see \cite{CGC-cry-art-BES}, \cite{CGC-cry-art-toli05}).
Generally speaking, the worst scenario consists of a space {\em large enough} to make AES linear but {\em small enough} to allow practical computations.
This is probably not possible.
Our goal is to find a space {\em small enough} to allow practical computations but {\em large enough} to identify a specific behavior of AES, showing that it is closer to linear maps than expected.  

In Section 1, after some basic algebraic background, we explain our point of view on {\em block ciphers}. 
In particular, we introduce the class of {\em translation based} cryptosystems, which are ciphers enjoying  some interesting algebraic properties.
We also briefly describe the three main translation-based cryptosystems: AES, SERPENT and PRESENT.

For completeness, in Section 2 we list the best-known attacks on round-reduced versions of AES.

In Section 3  we provide formal techniques to construct a larger space on which the block cipher can act. We call these techniques {\it space embeddings}.
In the case of translation-based ciphers, these embeddings are designed to lower the non-linearity of the encryption functions.
We present one specific embedding and we obtain several results on the rank distributions for matrices in the larger space, which are useful to mount attacks. 

In Section 4 we present a larger embedding, that apparently works well with AES and other translation-based systems. 
The effectiveness of this embedding depends heavily on properties of the mixing-layer.

In Section 5 we outline our approach to attack translation-based ciphers (including AES) with our embeddings. 
Although we have not been able to find an attack giving satisfactory statistical evidence, we have some partial data suggesting that our methods may work, as reported in \cite{CGC-cry-prep-BRS10} 

In Section 6 we discuss further on our non-linearity notion:
\begin{itemize}
\item first, we report results from \cite{CGC-tesi5-maines},\cite{CGC-alg-art-maines} on embeddings where the decrease in non-linearity can be formally proved;
\item then, we propose alternative embeddings highlighting their flaws;
\item finally, with group theory proofs we also show that it is very unlikely that a representation/embedding can completely linearize any version of AES.
\end{itemize}

\pagebreak
\section{Preliminaries}
In this section we recall well-known results in group theory and finite field theory \cite{CGC-cd-book-niederreiter97} 
in order to fix the notation we will use in the sequel. We also outline some basic ideas about {\em block ciphers} 
and we recall the structure of three well-known cryptosystems: AES, SERPENT and PRESENT.

\subsection{Group representations}
\label{Groups}
Let $n \geq 2$ be an integer. Let $V=(\FF_2)^n$ be the vector space over the finite field $\FF_2$ of dimension $n$.
We denote by $\Sym(V)$ and $\Alt(V)$, respectively, the symmetric and 
alternating group on $V$.
For any $N$,  we denote by $\Sym_N$ and $\Alt_N$, respectively, the symmetric and 
alternating group on $\{1, \ldots, N\}$. Clearly $\Sym(V)$ is isomorphic to $\Sym_{2^n}$ (the same for the alternating group). 
We denote by $\GL(V)$ the group of all linear permutations of $V$.
We recall the well-known formulas:
$$
  |\Sym(V)|= 2^{n}!, \quad |\Alt(V)|=\frac{2^{n}!}{2} \quad
  |\GL(V)|= \prod_{h=0}^{n-1}(2^{n}-2^h)<2^{n^2} \,.
$$
Given a finite group $G$, we say that $G$ can be {\sf linearized} if there is an injective morphism $\rho:G \rightarrow \GL(V)$
(this is called a ``faithful representation'' in representation theory).
If $G$ can be linearized, then, for any element $g \in G$, we can compute a matrix $M_g$ corresponding to the action of 
$g$ over $V$ (via $\rho$). The matrix computation is easy, since it is enough to evaluate $g$ on a basis of $V$.\\
If $\rho : G \rightarrow \GL(V)$ is a representation of $G$ on $V$, then we often write $vg$
instead of $v\rho(g)$, if no confusion arises. Also, $G$ is said to {\em act linearly} on $V$,
and $V$ is called a $G$-{\em module}. The {\em degree} of the representation is by definition
the dimension of $V$.
If we consider $\Sym_{N}$, we can always linearize $\Sym_{N}$ over $V$ via the so-called {\em regular} representation as follows.\\
Let $V$  be a vector space with basis $\{e_{1} ,\ldots ,e_{N}\}$.
The {\sf regular} representation $\rho: \Sym_{N} \rightarrow \GL(V)$ is defined by 
$(e_i)\rho(g) = e_{ig}$.
In other words, any permutation in $\Sym_{N}$ is associated to a permutation $(N \times N)$ matrix (and viceversa).
Since any finite group $G$ can be embedded in $\Sym_{N}$ for a smallest $N$, we can always linearize $G$ using the regular representation. 
But of course this is huge and usually impractical.

\subsection{Finite Fields}
\label{prel-ff}
For any prime $p$ and any positive $m\in \NN$, $\FF_{p^m}$ is the field with $p^m$ elements (unique up to field isomorphism).   
It contains an isomorphic copy of $\FF_p$ and can thus be thought as an extension of $\FF_p$.
On the other hand, we can construct any $\FF_{q^s}$ from $\FF_q$ with $q=p^m$ elements, as follows.

\pagebreak
Let $f \in \FF_q[x]$ be an irreducible polynomial of degree $m$. We can consider the quotient $R=\FF_q[x]/(f)$, 
where $(f)$ is the ideal generated by $f$ in $\FF_q[x]$. 
By considering the natural projection $\pi:\FF_q[x] \rightarrow R$, we call $\alpha=\pi(x)$ and clearly any element of $R$ 
can be uniquely expressed as a polynomial in $\alpha$ of degree less than $m$:
$$
R=\left\{\sum_{i=0}^{m-1}a_i \alpha^i \mid a_i \in \FF_q\right\}
$$ 
with the condition $f(\alpha)=0$.
\begin{theorem}
$R=\mathbb{F}_{q}[x]/(f)$ is a field and $R \cong \FF_{q^m}$. 
\end{theorem}
We denote by $\FF_q^*$ the multiplicative group of non-zero elements of $\FF_q$.
\begin{theorem}
For any finite field $\FF_q$, the multiplicative group $\FF_q^*$ is cyclic.
\end{theorem}
A generator of the cyclic group  $\FF_q^*$ is called a {\em primitive element} of $\FF_q$.

\begin{definition}
An irreducible polynomial $f \in \FF_q[x]$ is {\sf primitive} if its roots are primitive elements.
\end{definition}
Note that for any $q$ and $m$ there are indeed irreducible polynomials of degree $m$ over $\FF_q$ and some of them are primitive.
\subsection{Permutation polynomials}
\begin{definition}
A polynomial $f \in \FF_q[x]$ is a {\sf permutation polynomial} of $\FF_q$ if the associated polynomial function 
$f:c \mapsto f(c)$ from $\FF_q$ into $\FF_q$ is a permutation of $\FF_q$.
If $f$ is an affine map $f:x \mapsto ax+b$ $(a \not=0)$, we say that $f$ is a {\sf linear polynomial}.
\end{definition}

We note the following easy results:
\begin{enumerate}
\item Every linear polynomial over $\FF_q$ is a permutation polynomial of $\FF_q$.
\item The monomial $x^n$ is a permutation polynomial of $\FF_q$ if and only if $$\gcd(n,q-1)=1.$$
\end{enumerate}
Permutation polynomials of $\FF_q$ of degree less then $q$ can be combined by the operation of composition and 
subsequent reduction modulo $x^q-x$. The set of permutation polynomials of $\FF_q$ of degree less then $q$ forms a group, 
which is isomorphic to $\Sym(\FF_q)$. Then, the symmetric group $\Sym(\FF_q)$ and its subgroups can be represented as 
groups of permutation polynomials.
\begin{theorem}
\label{PermutPolyTheor}
For $q > 2$, the symmetric group $\Sym(\FF_q)$ is generated by $x^{q-2}$ and all linear polynomials over $\FF_q$.
\end{theorem}

\subsection{Block ciphers}
\label{Block Ciphers}
Block ciphers form an important class of cryptosystems in symmetric key cryptography. 
These are algorithms that encrypt and decrypt blocks of data (with fixed length\footnote{Actually, 
there is a recent approach that allows a slight change of the block length \cite{CGC-cry-art-CookYK09}}) 
according to a shared secret key. 
We can formally describe such a cryptosystem using the following definition: 
\begin{definition}
A cryptosystem is a pair $(\mathcal{M},\mathcal{K})$, where:
\begin{itemize}
\item $\mathcal{M}$ is a finite set of possible messages (plaintexts, ciphertexts);
\item $\mathcal{K}$, the key-space, is a finite set of possible keys;
\item we have encryption and decryption functions for any key
$k\in \mathcal{K}$:
$$
\phi_k: \mathcal{M}\rightarrow \mathcal{M}, \quad \psi_k: \mathcal{M} \rightarrow \mathcal{M}, \quad \phi_k, \psi_k \in \Sym(\mathcal{M})
$$
such that 
\begin{equation*}
 \psi_k = (\phi_k)^{-1}.
\end{equation*}
\end{itemize}
\end{definition}

Following the most used structure in modern ciphers, in the previous definition we set that the plaintext space
coincides with the ciphertext space.
W.l.o.g, we can consider  $\mathcal{M}=(\FF_q)^r$ and $\mathcal{K}=(\FF_q)^{\ell}$, 
with $r$ and $\ell$ positive integers, and we change slightly our previous definition.

\begin{definition}
Let $r$ and $\ell$ be natural numbers. Let $\phi$ be any function
$$
\phi: (\FF_q)^r\times (\FF_q)^{\ell} \rightarrow (\FF_q)^{r}.
$$
For any $k\in (\FF_q)^{\ell}$, we denote by $\phi_k$ the function
$$
 \phi_k:  (\FF_q)^r \rightarrow (\FF_q)^r, \quad \phi_k(x)=\phi(x,k).
$$
We say that $\phi$ is a {\bf algebraic} {\sf block cipher} if $\phi_k$ is a
permutation of $(\FF_q)^r$ for any key $k\in (\FF_q)^{\ell}$.
\end{definition}
\noindent
Under this conditions, we can also consider a block cipher as an indexed set of permutations
$(\FF_q)^{\ell} \rightarrow \Sym((\FF_q)^r).$
Any key $k\in \mathcal{K}$ induces a permutation $\phi_k$ on $\MM$.
Since $\MM$ is usually $V=(\FF_2)^r$ for some $r\in \mathbb{N}$, we can consider
$\phi_k\in \Sym(V)$.\\

To achieve the desired security, most modern block ciphers are {\sf iterated ciphers} that typically incorporate 
sequences of permutation and substitution operations. 
In fact, according to the ideas that Shannon proposed in his seminal paper \cite{CGC-cry-art-shannon}, 
the encryption process takes as input a plaintext and a random key and so proceeds through $N$ similar rounds.
In each round (except possibly for a couple, which may be slightly different) the iterated ciphers perform 
a non-linear substitution operation (or {\sf $S$-box}) on disjoint parts of the input that provides ``confusion'', 
followed by a permutation (usually a linear/affine transformation) on the whole data that provides ``diffusion''.
A cryptosystem reaches ``confusion'' if the relationship between plaintext, ciphertext and key is very complicated. 
The ``diffusion'' idea consists of spreading the influence of all parts of the input (plaintext and key) 
to all parts of the ciphertext. The operations performed in a round form the
{\sf round function}.
The round function at the $\rho$-th round ($1 \leq \rho\leq N$) takes as inputs both the output of the $(\rho-1)$-th round 
and the subkey $k^{(\rho)}$ (also called {\sf round-key}). Any round key $k^{(\rho)}$ is constructed starting from a 
{\sf master key}\footnote{also called {\sf session key}.} $k$ of some specified length, e.g. $k\in \mathcal{ K}=(\FF_2)^{\ell}$ 
(nowadays we have $2^{64}\leq|\mathcal {K}|\leq 2^{256}$). The {\sf key schedule} is 
a public algorithm (strictly dependent on the cipher) which constructs $N+1$ subkeys  $(k^{(0)}, \ldots, k^{(N)})$.\\

Several independent formal definitions have been proposed for iterated block ciphers (or subclasses of them).
Stinson in \cite{CGC-cry-book-stin95} gives the following definition of {\em substitution permutation network} (SPN for short).
In \cite{CGC-cry-book-aesbook} we can find another class of iterated block cipher, called the {\em key-alternating} block ciphers.

Now, we consider a more recent definition \cite{CGC-cry-art-carantisalaImp} that defines a class 
(see Definition \ref{tb definition}), large enough to include some common ciphers, 
yet restricted enough to have simple criteria guaranteeing an interesting property of the cipher 
(for details see Subsection \ref{Further remark}).

Let $V=(\FF_2)^r$ with $r=m b$, $b\geq 2$. The vector space $V$ is a direct sum
\begin{equation*}
  V = V_{1} \oplus \dots \oplus V_{b},
\end{equation*}
where each $V_{i}$ has the same dimension $m$ (over $\FF_2$). 
For any $v\in V$, we will write $v = v_{1} \oplus \dots \oplus v_{b}$, 
where $v_{i} \in V_{i}$. 
Also, we consider the projections $\pi_{i} : V \to V_{i}$ mapping
$v \mapsto v_{i}$.

Any $\gamma\in \Sym(V)$ that acts as $v \gamma = v_{1} \gamma_{1} \oplus \dots \oplus v_{b} \gamma_{b}$,
for some $\gamma_{i}\in \Sym(V_i)$, is a {\sf bricklayer transformation} (a ``parallel map'')
and any $\gamma_i$ is a {\sf brick}.
The maps $\gamma_i$'s are traditionally called $S$-boxes and map $\gamma$
is called a ``parallel S-box''.
A linear (or affine) map $\lambda:V\rightarrow V$ is traditionally called a ``Mixing Layer''
when used in composition with parallel maps.
We denote by $\sigma_v$ a translation over $V$.
\begin{definition}
 A linear map $\lambda\in \GL(V)$  is a {\sf proper mixing layer} if
no sum of  some  of the  $V_{i}$  (except $\{0\}$  and $V$)  is
    invariant under $\lambda$. 
\end{definition}    

We can characterize the ``translation based'' class by the following

\begin{definition} 
\label{tb definition}
We say that ${\mathcal C}$ is 
{\sf translation based (tb)} if:
\begin{itemize} 
\item it is the composition of a finite number of rounds, 
      such that any round $\tau_{k}$ can be 
      written\footnote{we drop round indexes.} as 
      $\gamma\lambda\sigma_{\bar k}$, where
      \begin{itemize}
      \item $\gamma$ is a round-dependent bricklayer transformation
            (but it does not depend on $k$),
      \item $\lambda$ is a round-dependent linear map
            (but it does not depend on $k$),
      \item $\bar{k}$ is in $V$ and depends on both $k$ and the round
            ($\bar{k}$ is called a ``round key'');
      \end{itemize}
\item for at least one round we have (at the same time)
      that $\lambda$ is proper and
      that the map $\mathcal{K}\rightarrow V$, $k\mapsto \bar{k}$, is surjective
      (a ``proper'' round).
\end{itemize}
\end{definition}

In \cite{CGC-cry-art-carantisalaImp} the authors gave several non-trivial remarks that can be useful. 
Let us recall the principal ones.  

\begin{remark} \label{notrivial}
A generalization is obtained by allowing a key-independent permutation at the beginning and/or another at the end.
This is the case for example for the SERPENT cipher. Since these permutations have no influence on the cryptanalysis of a cipher, 
they can be ignored.
\end{remark}

\begin{remark}\label{gamma0}
A round consisting of only a translation is still acceptable, by assuming $\gamma=\lambda=1_V$ (the identity map on $V$), 
although obviously it is not proper. Indeed, we can always assume that the first round is of this kind, 
otherwise we can remove its $\gamma$ and $\lambda$ (Remark \ref{notrivial}). 
Then, we can also assume that $0\gamma=0$, since we can add $0\gamma$ to the round key of the previous round.\\ 
If the previous round is proper, it remains proper since $\sigma_{0\gamma}$ is a permutation over $V$.
\end{remark}

\begin{remark}
To allow affine mixing layers,  rather than linear mixing layers, seems a generalization.
 However, this case is indeed already present
in Definition \ref{tb definition}, since it is enough to change $\sigma_v$ to incorporate the
``translation part'' of the mixing layer.
\end{remark}

\begin{remark}
A generalization can be obtained by only requiring \emph{at least one} of the
rounds to be of the prescribed form (with a proper mixing layer).
Although the authors' results still hold in this more general case,
we do not know any interesting cipher of this kind.
\end{remark}

Note that some famous ciphers, such as the DES, KASUMI and IDEA ciphers,
cannot be seen easily as {\sf tb} ciphers. Some of them (e.g. DES and KASUMI) are of {\em Feistel} type.
They modify only one half of the cipher state in each round.
It has been suggested that the Feistel ciphers suffer from a slow speed of diffusion compared to SPN (or {\em key-iterated}) ciphers.  

In the Subsections \ref{AES description}, \ref{serpent-description}, \ref{present description} we are going to describe 
respectively AES, SERPENT and PRESENT as translation based cryptosystems\footnote{The reader can find a full description 
of these cryptosystems respectively in \cite{CGC-cry-book-aesbook}, \cite{CGC-cry-art-serpent} and \cite{CGC-cry-art-PRESENT} }.

\subsection{The AES-128 cryptosystem} 
\label{AES description}

Let $\MM =\mathcal{K}=V=(\FF_2)^r$ with $r=128$ and let $x \in \MM$ be our plaintext, 
$k \in \mathcal{K}$ our random key and $y=\phi_{k}(x)$ the corresponding ciphertext.
Before describing the individual components $\gamma$, $\lambda$ and $\sigma_{k}$ of the {\em round function},
we recall (see Section \ref{prel-ff}) that it is possible to identify $(\FF_2)^8$ with the field $\FF_{2^8}$, 
via the quotient map $\FF_{2^8} \leftrightarrow \FF_2 [x]/\langle {\sf m} \rangle$, 
where ${\sf m} \in \FF_2[x]$ is an irreducible polynomial such that $\deg({\sf m})=8$.
The irreducible (but not primitive) AES polynomial is ${\sf m}=x^8+x^4+x^3+x+1$.

Internally, the AES algorithm's operations are performed on a two-di\-men\-sional array of bytes, called the {\sf State}. 
It consists of $4$ rows and $4$ columns and each element of this matrix is one byte (i.e. an element of $\FF_{2^8}=\FF_{256}$).
At the start of the encryption process, 
the input ${\sf x}$ (the plaintext) is a vector in $V$ and it is first changed into a $16$-byte vector:
$$
  \nu: (\FF_2)^{128} \rightarrow (\FF_{256})^{16},\quad {\sf x}\mapsto {\sf y} .
$$
Each round performs its operations on the State and after the last round the State is ``unwrapped'' 
and ``fills up'' the output vector.\\

A preliminary translation $\sigma_{k^{(0)}}$, where $k^{(0)} \in (\FF_2)^{r}$ is the first round key, is applied to the plaintext to form the input to the ({\tt Round 1}). It means that we can consider a preliminary round ({\tt Round 0}) such that $\gamma=1_{V}$ and $\lambda=1_{V}$ (see Remark \ref{gamma0}).\\
In order to obtain the ciphertext, other $N=10$ rounds follow.\\ 
Let $1\leq \rho \leq N-1$. 
A typical round ({\tt Round $\rho$}) can be written as 
the composition\footnote{Note that the order of the operation is exactly: $\gamma$, $\lambda$, and then $\sigma_k$.}
$\gamma \lambda \sigma_{k^{(\rho)}}$, where
\begin{itemize}
\item the parallel map $\gamma$ is called {\sf SubBytes} and it works in parallel to each of the $16$ bytes of the data;
\item the affine map $\lambda$ is the composition of two operations known as  {\sf ShiftRows} and {\sf MixColumns};
\item $\sigma_{k^{(\rho)}}$ is the translation with the session key $k^{(\rho)}$ (this operation is called \\{\sf AddRoundKey}).
\end{itemize}

The last round ({\tt Round $N$}) is atypical and is characterized by $\gamma \bar{\lambda} \sigma_{k^{(N)}}$ where
the affine map $\bar{\lambda}$ is only made by the {\sf ShiftRows} operation. 
So we obtain our ciphertext ${\sf y}=\phi_k({\sf x})$.\\
In the following, we analyze the structure of each component of the round function.
\pagebreak
\subsubsection{SubBytes}
\label{SubBytes}
The vector space $V$ is the direct sum $V=V_{1} \oplus \cdots \oplus V_{16}$ where each $V_i=(\FF_2)^8$ ($1\leq i\leq 16$). 
Any parallel map $\gamma \in \Sym(V)$ acts on an element $v \in V$ as  $v\gamma=v_1\gamma_1 \oplus \ldots \oplus v_{16}\gamma_{16}$, where $v_{i} \in V_i$ and $\gamma_i \in \Sym(V_i)$.
The {\sf SubBytes} operation $\gamma$ is composed by two transformations: the inversion in $\FF_{2^8}$ and an affine transformation.\\
The {\em inversion operation} is the {\em patched inversion}\footnote{Since the AES consists of $10$ rounds and each round requires $16$ $S$-box computations, the probability of there being no $0$-inversions during an encryption is $(255/256)^{160}\approx 0.53$.} in $\FF_{2^8}$ (i.e. $\varphi(x)=x^{254}$).\\
The {\em affine transformation} over $\FF_{2}$ consists of an affine mapping 
$\xi : (\FF_{2})^8 \rightarrow (\FF_2)^8$, 
specified by an $8 \times 8$ circulant matrix over $\FF_2$ and a translation. The result of inversion is regarded as a vector in $(\FF_2)^8$ and the output is given by $y=\xi(x)$, where
\begin{displaymath}
\left( \begin{array}{c}
y_7\\
y_6\\
y_5\\
y_4\\
y_3\\
y_2\\
y_1\\
y_0\\
\end{array} \right)
 = 
\left( \begin{array}{cccccccc}
1 & 0 & 0 & 0 & 1 & 1 & 1 & 1\\
1 & 1 & 0 & 0 & 0 & 1 & 1 & 1\\ 
1 & 1 & 1 & 0 & 0 & 0 & 1 & 1\\ 
1 & 1 & 1 & 1 & 0 & 0 & 0 & 1\\
1 & 1 & 1 & 1 & 1 & 0 & 0 & 0\\
0 & 1 & 1 & 1 & 1 & 1 & 0 & 0\\   
0 & 0 & 1 & 1 & 1 & 1 & 1 & 0\\
0 & 0 & 0 & 1 & 1 & 1 & 1 & 1\\  
\end{array} \right)
\left( \begin{array}{c}
x_7\\
x_6\\
x_5\\
x_4\\
x_3\\
x_2\\
x_1\\
x_0\\
\end{array} \right)
+
\left( \begin{array}{c}
0\\
1\\
1\\
0\\
0\\
0\\
1\\
1\\
\end{array} \right)
\end{displaymath}

\subsubsection{Mixing Layer}
\label{MixingLayer}
The map $\lambda:V \rightarrow V$ is a composition of two linear operations: {\sf ShiftRows} and {\sf MixColumns}.
The {\sf ShiftRows} operation is performed as follows. 
Any byte (an element of $\FF_{2^8}$) in row $i$ of the State, where $0 \leq i \leq 3$, is cyclically shifted (towards
left) by $i$ positions, as follows:
\ \\
$$
\begin{array}{|c|c|c|c|}
\hline
s_{0}& s_{4}& s_{8}& s_{12} \\
\hline
s_{1}& s_{5}& s_{9}& s_{13} \\
\hline
s_{2}& s_{6}& s_{10}& s_{14} \\
\hline
s_{3}& s_{7}& s_{11}& s_{15} \\
\hline
\end{array}
\quad \rightarrow \quad \textrm{ShiftRows} \quad \rightarrow \quad
\begin{array}{|c|c|c|c|}
\hline
s_{0}& s_{4}& s_{8}& s_{12} \\
\hline
s_{5}& s_{9}& s_{13}& s_{1} \\
\hline
s_{10}& s_{14}& s_{2}& s_{6} \\
\hline
s_{15}& s_{3}& s_{7}& s_{11} \\
\hline
\end{array}
$$
\ \\

\pagebreak
In other words, we can describe the {\sf ShiftRows} operation by the map $${\sf sh}:(\FF_{2^8})^{16} \rightarrow  (\FF_{2^8})^{16}$$
$$(s_0,s_1,\cdots,s_{15})  \mapsto (s_0,s_5,s_{10},s_{15},s_4,s_9,s_{14},s_3,s_{8},s_{13},s_2,s_{7},s_{12},s_{1},s_{6},s_{11}).$$ 

We can also represent the ShiftRows operation with the following $16 \times 16$ block diagonal matrix 
$$
S=\left( \begin{array}{cccc}
I &0 &0 &0\\
0& R &0 &0\\
0 &0& R^2 &0 \\
0& 0 &0& R^3 \\
\end{array} \right)
\quad  \quad
R=\left( \begin{array}{cccc}
0& 1 &0 &0\\
0 &0& 1 &0\\
0& 0 &0& 1\\
1 &0 &0 &0\\
\end{array} \right)
$$
where the matrix $R$ is a permutation matrix over $\FF_{2^8}$ that represents the shift of one row by one position.\\

In order to describe the {\sf MixColumns} operation, each column of the State can be treated as a four-term polynomial in $\FF_{256}[z]$.
Let $c(z)$ be one such polynomial. Then each column is replaced by the result of the multiplication in $\FF_{256}[z]/(z^4+1)$ by $a(z)$,
$ c \mapsto c\cdot a\; \mod(z^4+1)$,
$$
\left( c_1,c_2,c_3,c_4 \right) \longrightarrow 
\left( c_1\cdot a, c_2 \cdot a, c_3\cdot a, c_3\cdot a \right).
$$
Note that $a(z)$ is invertible in $\FF_{256}[z]/(z^4+1)$. On the other hand, we can see the {\sf MixColumns} operation as a $4$-block diagonal matrix, each block the same MDS matrix (i.e. all minors are non-zero):

\begin{displaymath}
\left( \begin{array}{cccc}
z   &z+1 &1   &1\\
1   &z   &z+1 &1\\
1   &1   &z   &z+1 \\
z+1 &1   &1   &z \\
\end{array} \right)
\end{displaymath}

\begin{remark} This MDS property is used to ensure that the number of active S-boxes involved in a differential or linear attack increases rapidly, and the security of the AES against these particular attacks can be established.
\end{remark}

Obviously, we can also see the whole Mixing Layer ($\lambda$ linear operation) as a matrix ${\bf M}$. We observe that the order of this matrix is quite small, i.e. ${\bf M}^8=1$. (Also, both the order of {\sf ShiftRows} and {\sf MixColumns} are equal to $4$.)

\pagebreak
\subsection{The SERPENT cryptosystem}
\label{serpent-description}
Let $\MM=V=(\FF_2)^r$, with $r=128$. We consider $\mathcal{K}=(\FF_2)^{\ell}$, with the fixed length $\ell=128$, 
although the key is designed with variable length.\\
The encryption $\phi$ proceeds by $N=32$ similar rounds and it works as follows:
\begin{itemize}
\item a preliminary permutation is applied $\pi: V \rightarrow V$
(this is not used for security, rather to ease the implementation);
\item there is a preliminary translation with the first round key; 
\item $N-1$ rounds with the same structure are applied, but using a different permutation, 
each composed of a key translation $\sigma_{k}$, a parallel S-box $\gamma$ and a linear mixing-layer $\lambda$
(we denote the round $\rho$ by {\tt Round $\rho$}, with $ \rho=1,...,31$);
\item the last round ({\tt Round $32$}) follows and it consists of the composition $\gamma \lambda \sigma_k$ where $\lambda=1_V$;
\item a final permutation $\pi^{-1}: V \rightarrow V$ is performed.
\end{itemize}

The decryption process is easily obtained by inverting every step of the encryption,
using the inverse of the $S$-boxes, the inverse of the mixing-layer and
the reverse order of the round keys.\\

Let $\rho$ be a natural number such that $1\leq \rho\leq 31$.
In order to describe a typical round ({\tt Round $\rho$}) we have to specify how the components 
$\gamma$, $\lambda$ and  $\sigma_k$ are applied.
We note that, after the permutation $\pi:V\rightarrow V$, 
we perform a preliminary translation $\sigma_{k^{(0)}}$, where $k^{(0)} \in (\FF_2)^{r}$ is the first round key.

Let $V=V_1 \oplus \cdots \oplus V_{32}$, where , for any $1\leq j\leq 32$,  each $V_{j}=(\FF_2)^4$. 
Any $\gamma \in \Sym(V)$ acts as  $v\gamma=v_1\gamma_1 \oplus \ldots \oplus v_{32}\gamma_{32}$, 
where $v_{j} \in V_j$ and $\gamma_j \in \Sym(V_j)$.

We have to characterize each $\gamma_j$ (i.e. we have to construct each $S$-box).
The eight $S$-boxes $S_1,\ldots, S_8$ of SERPENT were built ``ad hoc'' starting from the $8$ fixed $S$-boxes of DES (see \cite{CGC-cry-art-serpent}).
To each $v_j$ we apply the  same $S$-box  $S_{i \mod 8}$, so that $S_{i \mod 8}(v_j)$ lies in $(\FF_2)^4$.
That is, $\gamma_1=\gamma_2= \cdots =\gamma_{32}=S_{i \mod 8}$.

Then the linear transformation $\lambda$ (described in \cite{CGC-cry-art-serpent}) and a final translation $\sigma_{k^{(\rho)}}$ are applied.
The last round ({\tt Round $32$}) is only slightly different. 
The only difference with a typical round is the replacing of the linear transformation $\lambda$ by $1_V$.

\pagebreak
\subsection{PRESENT: an ultra-lightweight block cipher}
\label{present description}
PRESENT is an iterated block cipher that consists of $N=31$ rounds.\\
Let $\MM =V=(\FF_{2})^r$ with $r=64$. Let $\mathcal{K}=(\FF_{2})^{\ell}$, where $\ell$ may be equal to $80$ or $128$.
We consider only the PRESENT's version such that $\mathcal{K}=(\FF_{2})^{80}$, 
since its authors recommend it in order to have a good performance.\\ 
We are going to describe how the {\em round function} $\gamma \lambda \sigma_{k^{(\rho)}}$ (in the $\rho$-th typical round) 
is performed.\\
As in the AES and SERPENT cryptosystems, the encryption process starts with a preliminary round ({\tt Round 0}) 
that consists of a parallel map $\gamma=1_V$, a linear transformation $\lambda=1_V$ and the translation $\sigma_{k^{(0)}}$ , 
where $k^{(0)} \in (\FF_2)^{r}$ is the first round key.
A typical round consists of the non-linear operation, called {\sf sBoxLayer}, 
the linear transformation, known as {\sf pLayer} and the sum with the round key.\\   

The parallel map $\gamma \in \Sym(V)$ used in PRESENT acts as $$v\gamma=v_1\gamma_1 \oplus \ldots \oplus v_{16}\gamma_{16},$$ 
where each 
$v_{i} \in (\FF_{2})^{4}$ and $\gamma_i \in \Sym((\FF_{2})^{4})$  ($1\leq i \leq 16$). 
The action of any brick $\gamma_i : (\FF_2)^4 \rightarrow(\FF_2)^4$ is given by the following table, using an hexadecimal notation:

$$
\begin{array}{|c||c|c|c|c|c|c|c|c|c|c|c|c|c|c|c|c|c|}
\hline 
         x &0 &1 &2 &3 &4 &5 &6 &7 &8 &9 &A &B &C &D &E &F\\
 \hline        
 \gamma[x] &C &5 &6 &B &9 &0 &A &D &3 &E &F &8 &4 &7 &1 &2\\
\hline 
\end{array}
$$

\ \\

The affine map $\lambda:V \rightarrow V$ is a bit permutation as given by the following table,
where the bit $i$ of the intermediate state is moved to the bit position $P(i)$.

$$
\begin{array} {|c c c c c c c c c c c c c c c c c|}
\hline
i &0 &1 &2 &3 &4 &5 &6 &7 &8 &9 &10 &11 &12 &13 &14 &15\\
P(i) &0 &16 &32 &48 &1 &17 &33 &49 &2 &18 &34 &50 &3 &19 &35 &51\\
\hline
\hline
i &16 &17 &18 &19 &20 &21 &22 &23 &24 &25 &26 &27 &28 &29 &30 &31\\
P(i) &4 &20 &36 &52 &5 &21 &37 &53 &6 &22 &38 &54 &7 &23 &39 &55\\
\hline
\hline
i &32 &33 &34 &35 &36 &37 &38 &39 &40 &41 &42 &43 &44 &45 &46 &47\\
P(i) &8 &24 &40 &56 &9 &25 &41 &57 &10 &26 &42 &58 &11 &27 &43 &59\\
\hline
\hline
i &48 &49 &50 &51 &52 &53 &54 &55 &56 &57 &58 &59 &60 &61 &62 &63\\
P(i) &12 &28 &44 &60 &13 &29 &45 &61 &14 &30 &46 &62 &15 &31 &47 &63\\
\hline
\end{array}
$$

%
\newpage
\section{Known attacks}
\label{structural AES}
AES's structure has been used to carry out some innovative analysis.
Such attacks tend to have a similar form:
\begin{itemize}
\item they identify a property holding for a few rounds with a good probability;
\item they use special techniques to extend the attack to more rounds. 
\end{itemize}

The following table summarizes the more successful attacks on round-reduced versions of the AES cryptosystem:
$$
\begin{small}
\begin{tabular}{|c||c c c l l|}
\hline
Key&\textrm{Rounds}& \textrm{Texts}& \textrm{Time}& \textrm{Type}& \textrm{Reference}\\
\hline
\hline
128 & $5$ & $2^{11}$            & $2^{40}$  & Square attack    & \cite{CGC-cry-art-deamenrijmen1}\\
128 & $5$ & $2^{29.5}$          & $2^{31}$  & Impossible diff. &  \cite{CGC-cry-art-bihamkeller}\\
128 & $5$ & $2^{39}$            & $2^{39}$  & Boomerang attack &  \cite{CGC-cry-art-biryukov}\\
128 & $6$ & $2^{32}$            & $2^{72}$  & Square attack    &  \cite{CGC-cry-art-deamenrijmen1}\\
128 & $6$ & $2^{34.6}$          & $2^{44}$  & {\sf Partial Sum}      & \cite{CGC-cry-art-ferguson1}\\
128 & $6$ & $2^{91.5}$          & $2^{122}$ & Impossible diff. & \cite{CGC-cry-art-cheon}\\
128 & $6$ & $2^{71}$            & $2^{71}$  & Boomerang attack & \cite{CGC-cry-art-biryukov}\\
128 & $7$ & $2^{128}-2^{119}$   & $2^{120}$ & Partial Sum      & \cite{CGC-cry-art-ferguson1}\\
128 & $7$ & $2^{32}$            & $2^{128}$ & Collision        &  \cite{CGC-cry-art-gilbert}\\
\hline
192 & $7$ & $2^{91.2}$            & $2^{139.2}$ & Impossible diff. & \cite{CGC-cry-art-LuDunkelman08}\\
192 & $8$ & $2^{127}$            & $2^{188}$ & Partial Sum & \cite{CGC-cry-art-ferguson1}\\
192 &$10$ & $2^{124}$            & $2^{183}$ & (Related-key) Rectangle & \cite{CGC-cry-art-BihDunKel05a}\\
192 &$12$ & $2^{123} $          & $2^{176}$  & {\sf (Related-key) Ampl. Boomerang}                  &\cite{CGC-cry-prep-biryukov09} \\
\hline
256 & $8$ & $2^{32}$            & $2^{194}$ & Partial Sum & \cite{CGC-cry-art-ferguson1}\\
256 & $9$ & $2^{85}$            & $2^{126}$ & Partial Sum & \cite{CGC-cry-art-ferguson1}\\
256 &$10$ & $2^{114}  $         & $2^{173}$ & (Related-key) Rectangle  &\cite{CGC-cry-art-BihDunKel05a} \\
256 &$14$ & $2^{119}  $         & $2^{119}$ & {\sf (Related-key) Boomerang}  & \cite{CGC-cry-prep-biryukov09}\\
\hline 
\end{tabular}
\end{small}
$$
Other researchers attack small scale variants of the AES, where also the message space and the key space are reduced (see e.g.\cite{CGC-cd-inbook-D1cid}).
A recent practical attack (due to A.Biryukov, O.Dunkelman, N.Keller, D.Khovratovich, A.Shamir) on a ($10$-round version) of AES-$256$ has been presented (\cite{CGC-cry-art-BDKKS10}).

\section{First results}

In the literature there are some ways of representing the same cipher (e.g. AES), 
like BES \cite{CGC-cry-art-BES} or Dual Ciphers \cite{CGC-cry-art-barkanbiham}, that could be useful for the cryptanalysis. 
Other ways of representing AES that exploit its structure can be found, for example, in  \cite{CGC-cry-book-cidMurphyRobshaw}.\\	
In this section we represent ``AES-like'' ciphers by embedding them into larger ciphers. In Subsection \ref{preliminary rep} we begin with
We want to enlarge $\Omega$ to a set $W$ such that:
\begin{enumerate}
\item $W$ is endowed with a vector space structure;
\item  the permutations can be extended to act linearly on the whole $W$.
\end{enumerate}
In Subsection \ref{FirstRepresent} we provide one specific embedding of AES-like ciphers 
that linearizes the non-linear part of these ciphers, but it fails to linearize the whole cipher.
In particular our embedding can be applied to AES, PRESENT and SERPENT.

\subsection{Some preliminary results}
\label{preliminary rep}
Let $\Omega$ be a set such that $|\Omega|=n$, let $\Sym(\Omega)$ be the symmetric group on $\Omega$ and 
let $W$ be a vector space over a field $\FF$ (not necessarily a finite field).

\begin{definition}
\label{space embedding}
Let $G \leq \Sym(\Omega)$. An injective map $\phi: \Omega \rightarrow W$ is a {\bf space embedding} 
with respect to the group $G$ if,   
 $\forall \sigma  \in G$, $\exists A_{\sigma} \in \GL(W)$ such that $\phi \circ\sigma = A_{\sigma} \circ \phi .$
\end{definition} 

Moreover, $\phi(\Omega)$ is the set of all {\sf admissible vectors} (w.r.t. $\phi$),
the subspace $\langle\phi(\Omega)\rangle$ is the {\sf admissible space}.
Note that  since $\phi(\Omega) \subset \langle\phi(\Omega)\rangle$ 
then $\langle\phi(\Omega)\rangle$ is the smallest subspace containing all admissible vectors. 
Generally speaking, $|\langle \phi(\Omega)\rangle| >> |\phi(\Omega)|$.

Note that the {\sf regular representation} (see Subsection \ref{Groups}) can be considered as a {\sf space embedding} 
$\phi: \Omega \rightarrow W$ with respect to the  group $G = \Sym(\Omega)$, 
where $\dim(W)=|\Omega|=n$ and $\phi:\omega \mapsto b_{\omega}$ with $\{b_{\omega}\}_{\omega \in \Omega}$ a basis of $W$. 
Also, $W=\langle \phi(\Omega)\rangle$.

A space embedding permits to construct a faithful representation of $G$, as explained in the next proposition.
\begin{proposition}
\label{embedding-homomorphism}
Let $\alpha: \Omega \rightarrow W$ be a space embedding with respect to $G$. Suppose that 
$\forall \sigma \in G \quad \exists! A_{\sigma} \in \GL(W) \quad s.t. \quad \phi \circ \sigma = A_{\sigma} \circ \phi .$ 
Then
\begin{enumerate}
\item we can define a map $ \tilde{\phi}: G \rightarrow \GL(W)$, where $\tilde{\phi}(\sigma)=A_{\sigma}$, for any  $\sigma \in G$;
\item $ \tilde{\phi}$ is a group homomorphism.
\end{enumerate}
\pagebreak
\begin{proof}
$\emph{1}.$ Obvious.\\
$\emph{2}.$ We have to prove that $\tilde{\phi}(\sigma \sigma')=\tilde{\phi}(\sigma)\tilde{\phi}(\sigma')$ for all $\sigma,\sigma'\in G$, i.e. 
$A_{\sigma \sigma'}= A_{\sigma} A_{\sigma'}$.
Using Definition \ref{space embedding}, the following equality holds
$$A_{\sigma \sigma'}(\phi(\omega))=\phi((\sigma \sigma')(\omega))=\phi(\sigma(\sigma'(\omega))).$$
Since 
$$A_{\sigma}A_{\sigma'}(\phi(\omega))=A_{\sigma}(\phi(\sigma'(\omega)))=\phi(\sigma(\sigma'(\omega))),$$
we conclude that $A_{\sigma \sigma'}=A_{\sigma}A_{\sigma'}$, for all $\omega \in \Omega$.
\end{proof}
\end{proposition}
\begin{remark}
In Definition \ref{space embedding} we require only that $A_{\sigma}$ exists, however in Theorem \ref{embedding-homomorphism} we see that it is also unique.
\end{remark} 
For example, for the regular representation any permutation $\sigma \in \Sym(\Omega)$ defines a permutation $\sigma \in \Sym(\{b_{\omega}\}_{\omega \in \Omega})$ and so it defines a unique $A_{\sigma} \in \GL(W)$, which can be represented as a permutation matrix.

Now, we are interested in a special case of {\sf space embedding} where the set $\Omega$ is a vector space $V=(\FF_2)^{r}$ and  $W$ is the vector space $(\FF_2)^{s}$, with $s > r $. For any $1\leq i\leq s$, let ${\textbf e}_i\in W$:
$$
  {\textbf e}_i=(0,\dots,0,\underset{\underset{i}{\uparrow}}{1},0,\dots,0) \,.
$$
Let $\sigma \in \Sym(V)$ be any permutation over $(\FF_2)^{r}$. 
We want to embed $V$ into $W$ by an injective map $\alpha$ and to extend $\sigma$ to a permutation
$\sigma ' \in \Sym(W)$ as shown in the following commutative diagram:
$$
 \xymatrix{\ar @{} [dr] |{\circlearrowright}
   V \ar[d]^{\sigma} \ar[r]^{\alpha} & W \ar[d]^{\sigma'}\\
   V \ar[r]^{\alpha} & W}
$$

\noindent
In order to do this, we have to define the permutation $\sigma' \in \Sym(W)$. We say that $\sigma'$ is an {\sf extension} of $\sigma$. 
We seek a $\sigma'$ that is linear on $W$.
The following definition will be useful:
  
\begin{definition}
\label{linearly-extendible}
Let $\sigma \in \Sym(V)$ and $\alpha$ be an injective map $\alpha: V \rightarrow W$.
We say that $\sigma$ is {\bf linearly extendible} (via $\alpha$) if $\forall \{v^i\}_{i \in I} \subset V$ we have 
$$\sum_{i \in I}{\alpha(v^i)}=0 \iff \sum_{i \in I}{\alpha(\sigma(v^i))}=0.$$ 
\end{definition}

\begin{remark}
Since we are considering the finite field $\FF_2$, we note that $\sigma$ is linearly extendible (via $\alpha$) if $\forall \{v^i\}_{i \in I} \subset V$ such that $\sum_{i \in I}{\alpha(v^i)}=0$ we have $\sum_{i \in I}{\alpha(\sigma(v^i))}=0$. In fact, an injective map defined on the set
$$\{\{v^i\}_I\subset V \mid \sum_{i \in I}{\alpha(v^i)}=0\}\}$$ 
into the set 
$$\{\{\sigma(v^i)\}_I\subset V \mid \sum_{i \in I}{\alpha(\sigma(v^i))}=0\}\}$$ 
is a bijective map, since the cardinality of the two finite sets is the same.
\end{remark}

Let $\alpha:V\rightarrow W$ be a space embedding. Let $A=\mathrm{Im}(\alpha)=\alpha(V)$ and 
let $T=\langle A \rangle$ be the subspace (the admissible space) of $W$ linearly generated by $A$.
Since $\sigma '(\alpha(v))=\alpha(\sigma(v))$, $\forall v \in V$, we require that $\sigma'(A)=A$.

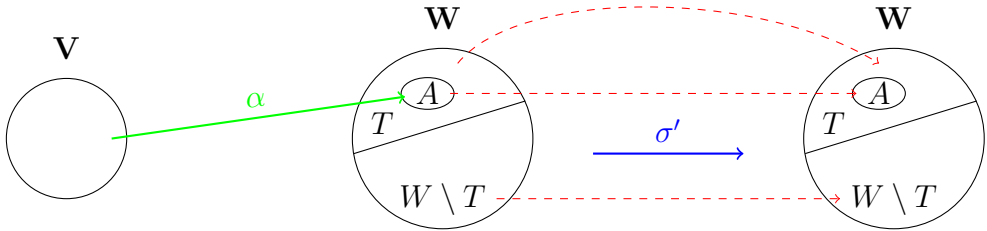
\begin{center}
\begin{tikzpicture}[scale=2]
\draw (-2.0,0.5) circle (.4cm) (0.5,0.5) circle (.6cm) (3.5,0.5) circle (.6cm);
\draw (-0.09,0.4) -- (1.05,0.75);
\draw (0.4,0.8) ellipse (5pt and 3pt);
\node (A1) at (0.4,0.8) {$A$};
\draw (3.4,0.8) ellipse (5pt and 3pt);
\node (A2) at (3.4,0.8) {$A$};
\draw (2.91,0.4) -- (4.05,0.75);
\node (T1) at (0.1,0.6) {$T$};
\node (T2) at (3.1,0.6) {$T$};
\node (WT1) at (0.5,0.1) {$W \setminus T$};
\node (WT2) at (3.5,0.1) {$W \setminus T$};
\node (W1) at (0.5,1.3) {$\mathbf{W}$};
\node (W2) at (3.5,1.3) {$\mathbf{W}$};
\node (V) at (-2.0,1.1) {$\mathbf{V}$};
\draw[red,->,dashed](A1) -- (A2);
\draw[red,->,dashed](WT1) -- (WT2);
\draw[green] [->, thick] (-1.7,0.5) -- node[above,sloped] {$\alpha$}(A1);
\draw[red,->,dashed](0.6,1.0) .. controls (1,1.5) and (2.6,1.5) .. (3.4,1.0);
\draw[blue] [->, thick] (1.5,0.4) -- node[above,sloped] {$\sigma '$}(2.5,0.4);
\end{tikzpicture}
\end{center}

In order to specify the behavior of $\sigma'$ on $(T\setminus A)$, which is the space of non-admissible vectors 
in the admissible space, we have to consider two different cases:
\begin{itemize}
\item[(a)] suppose that $\sigma$ is linearly extendible.
Let $t \in T$, we must have $t=\sum_{1\leq j \leq \iota}{a^j}$, with $\iota \geq 1$, with $\{a^j\}_{1\leq j\leq i} \subset A$, 
$a^j=\alpha(v^j)$ (with $1\leq j \leq \iota$ and $v^j \in V$).
Then we define $$\sigma'(t)=\sum_{1\leq j \leq \iota}{\sigma'(a^j)}=\sum_{1\leq j \leq \iota}{\alpha(\sigma(v^j))};$$ 
\item [(b)] in case $\sigma$ is not linearly extendible, we define $\sigma'_{\mid_{T\setminus A}}=\mathrm{id}_{T\setminus A}$.
\end{itemize}
We now define $\sigma '$ on $W \setminus T$ according to the two previous cases (i.e. depending on the behavior of $\sigma$ on A).
\ \\
In case (a), let $\tau$ be the dimension of the subspace $T$. We consider any subset $B$ of $\{{\textbf e}_1,\dots,
{\textbf e}_{s}\}$ such that $|B|=s-\tau$ and $W$ is the direct sum $W=T\oplus \langle B \rangle$. It is obvious that $B$ exists.
Let $w \in W$, then $w=w_T+w_B$ with $w_T \in T$ and $w_B \in \langle B \rangle $. Finally, we define 
$$\sigma '(w)=\sigma '(w_T)+ w_B .$$ 
In case (b) we define $\sigma'_{\mid_{W \setminus T}}= \mathrm{id}_{W\setminus T}$.

\begin{lemma}
\label{linear}
If $\sigma$ is linearly extendible, then $\sigma ' \in \GL(W)$.
\begin{proof}
We first show that $\sigma'$ is well-defined on $T$. 
Let $t=\sum_{I}{a^i}$ and $t'=\sum_{J}{a^j}$ and suppose that $t=t'$. Since $\sigma$ is linearly extendible, we have 
\begin{eqnarray*}
0=t+t'&=&\sum_{I}a^i+\sum_{J}a^j=\sum_{I}\alpha(v^i)+\sum_{J}\alpha(v^j)=\sum_{I\cup J}\alpha(v^i)\\
\sigma'(t)+\sigma'(t')&=&\sum_{I}{\sigma'(a^i)} +\sum_{J}{\sigma'(a^j)}=\sum_{I}\alpha(\sigma(v^i))+\sum_{J} \alpha(\sigma(v^j))\\
&=&\sum_{I \cup J}\alpha(\sigma(v^i))=0.
\end{eqnarray*}
We now show that $\sigma'$ is linear on $T$. 
Let $t_i=\sum_{h}a_h^{(i)}$. We have to show that $\sigma'(\sum_{i} t_i)=\sum_{i} \sigma'(t_i)$.
Clearly, 
 \begin{eqnarray*}
\sigma'\big(\sum_{i}\sum_{h}a_h^{(i)}\big)&=&\sigma'\Big(\sum_{i,h}a_h^{(i)}\Big)=\sum_{i}\sum_{h}\sigma'(a_h^{(i)})=
\sum_{i}\Big(\sum_{h}\sigma'(a_h^{(i)})\Big)\\
&=&\sum_{i}\sigma'(t_i)
\end{eqnarray*}
and we have our thesis.\\
Since $\sigma'$ is linear on $T$ and $T$ is a finite set, in order to prove that $\sigma'$ is bijective on $T$ it suffices 
to show that 
$\ker \sigma'=0$.
We have (by definition of linearly extendible)
\begin{eqnarray*}
0=\sigma'(t)= \sum \alpha(\sigma(v^j)) \iff 0=\sum \alpha (v^j)=t
\end{eqnarray*}  

Finally, we show the linearity on $W$.
Let $\{w^i\}_{i \in I} \subset W$, we have to show the following equality
\begin{equation}
\label{eqlinear}
\sigma' \Big(\sum_{i \in I} w^i \Big)= \sum_{i \in I} \sigma'(w^i).
\end{equation}
Since $W$ is direct sum of $T$ and $\langle B \rangle$,  each element $w^i$ in $W$ can be considered as $w^{i}_{T}+ w^{i}_{B}$ 
and so we can write the following
\begin{eqnarray*}
\sigma'\Big(\sum_{i \in I} w^i \Big) & = & \sigma'\Big(\sum_{i \in I} (w^{i}_{T}+w^{i}_{B}) \Big)  
= \sigma' \Big(\sum_{i \in I} w^{i}_{T} \Big) +\sum_{i \in I} w^{i}_{B}\\
\sum_{i \in I} \sigma'(w^i) & = & \sum_{i \in I} \sigma'(w^{i}_{T}+w^{i}_{B})   
=   \sum_{i \in I} \sigma'(w^{i}_{T}) +\sum_{i \in I} w^{i}_{B}.\\
\end{eqnarray*}
It easily follows that (\ref{eqlinear}) holds if and only if 
$$\sigma'\Big(\sum_{i \in I} w^{i}_{T}\Big)= \sum_{i \in I} \sigma'(w^{i}_{T}).$$  
\end{proof}
\end{lemma}
\begin{remark}
The construction of $\sigma' \in \GL(W)$ from $\sigma$ linearly extendible (Definition \ref{linearly-extendible})
can be done similarly over any field.\\
\end{remark}

We are now able to prove the main result of this subsection. 
\begin{theorem}
Let $W=(\FF_2)^r$ and $G \leq \Sym(V)$. An injective map $\alpha :V \rightarrow W$ is a space embedding with respect to $G$ 
if and only if, 
$\forall \sigma \in G$, $\sigma$ is linearly extendible.
\begin{proof}
Let $\alpha$ be a space embedding with respect to $G$. For any fixed $\sigma \in G$, 
there exists a map $A_{\sigma} \in \GL(W)$ such that $\alpha \circ \sigma=A_{\sigma} \circ \alpha$. 
Now, let $\{w^i\}_{i \in I}$ be a finite set such that $w^i=\alpha(v^i)$ (for any $i \in I$)
and $\sum_{i \in I}w^i=0$. Obviously we have
$$
\sum_{i \in I}\alpha(\sigma(v^i))=\sum_{i \in I} A_{\sigma}(\alpha(v^i))= 
\sum_{i \in I}A_{\sigma}(w^i)=A_{\sigma}\Big(\sum_{i \in I}w^i\Big)=0.
$$  
The converse immediately follows thanks to the previous lemma.
\end{proof}
\end{theorem}

\begin{remark}
For a fixed $\alpha$ and $\sigma$, the map $\sigma '$ is unique and $\tilde{\alpha} : G \rightarrow \GL(W)$
is a representation of $G$, by Proposition
\ref{embedding-homomorphism}.
\end{remark}
\begin{remark}
In the following we use $A_\sigma$ and $\sigma'$ interchangeably.
\end{remark}
\subsection{A first embedding}
\label{FirstRepresent}
We now apply the theory developed in the previous section to a specific 
space embedding\footnote{which is called ``$\alpha$'' in Subsection \ref{preliminary rep}.}
$\varepsilon:V \rightarrow W$.
 
Let us identify $(\FF_2)^m$ with the field $\FF_{2^m}$, via the quotient map 
$\FF_{2^m} \leftrightarrow \FF_2 [x]/\langle {\sf p} \rangle$, where 
${\sf p} \in \FF_2[x]$ is any primitive polynomial such that $\deg({\sf p})=m$. \\
We define a map $\varepsilon ': \FF_{2^m} \rightarrow (\FF_2)^{2^m}$ by means of a primitive element 
$\gamma$ of $\FF_{2^m}$ (which is a root of ${\sf p}$). 
The map $\varepsilon '$ is defined as 
$$
\varepsilon '(0)=(1,\underbrace{0,\dots,0}_{2^m -1}) \qquad 
\varepsilon '(\gamma^{i})=(0,\dots,0,\underset{\underset{i+1}{\uparrow}}{1},0,\dots,0) \quad \forall 1\leq i\leq 2^m-1 \,. 
$$ 
Note that $\varepsilon '(1)=\varepsilon '(\gamma^{2^m -1})=(\underbrace{0,\dots,0}_{2^m -1},1)$.\\
Let $b$ be a positive integer, let $r=m b$ and $s=2^m b$.  Let $V=(\FF_2)^r$ and $W=(\FF_2)^s$.
We construct our injective map $\varepsilon :V \rightarrow W$ in the following way:
\begin{equation}
\label{defepsilon}
\varepsilon(v_1,\dots,v_{b})=(\varepsilon '(v_1),\dots, \varepsilon '(v_{b}))
\end{equation}
for any $v_j \in (\FF_2)^m$ ( $1 \leq j \leq b$).
Note that $\varepsilon$ is a parallel\footnote{see Subsection \ref{Block Ciphers}.} map.\\

\noindent
For simplicity of notation, we set $e_1=\varepsilon '(0)=(1,\underbrace{0,\dots,0}_{2^m -1})$ and  
$e_{i+1}=\varepsilon '(\gamma^{i})$, for any $1\leq i\leq 2^m-1$.

\pagebreak
We note that
\begin{lemma}
\label{tecnico}
Suppose that $\sum_{i \in I}e_i=e_h.$ Then $h \in I$.
\begin{proof}
It follows from $\mathrm{w}(e_i)=1$, for all $i \in I$.
\end{proof}
\end{lemma}

The following lemma is easily proved:
\begin{lemma}
\label{parity}
Let $I$ be a finite index multiset such that $\{v^i\}_{I} \subset V$. 
For any $1\leq h\leq b$ we have $\sum_{i \in I}{\varepsilon '(v^i_h)}=0$ if and only if, 
$\forall i \in I$, $|\{j \in I \mid v^{j}_{h}=v^{i}_{h}\}|$ is even. 
\begin{proof}
Since  $\varepsilon'$ maps each element of  $(\FF_2)^m$  into the canonical basis of $(\FF_{2})^{2^m}$, 
each $ \varepsilon '(v^i_h)$ is a vector such that $\mathrm{w}(\varepsilon '(v^i_h))=1$.
Considering the following sum in $\FF_2$, we have that $\sum_{I}{\varepsilon '(v^i_h)}=0$ if and only if 
each component is made by an even number of $1$, i.e. 
if and only if each element of the canonic basis that appears in our sum has an even weight. 
Since $\varepsilon '$ is bijective, we have that $|\{j \in I \mid v^{j}_{h}=v^{i}_{h}\}|$ is even, $\forall i \in I$.
\end{proof}
\end{lemma}

\begin{proposition}
\label{Dimension}

Let $\varepsilon$ as in (\ref{defepsilon}). 
Then  $\dim_{\FF_2} \big(\langle \mathrm{Im}(\varepsilon) \rangle \big)= 2^m b -(b-1)$.
\begin{proof}
We  define the elements $$z_{i,j}=(e_1,\ldots,e_1,\underset{\underset{i}{\uparrow}}{e_j},e_1,\ldots,e_1),$$
for  $1\leq i \leq b$ and $1 \leq j \leq 2^m$. Note that $z_{i,j}\not= z_{h,\ell}$ for $(i,j)\not=(h,\ell)$, 
except for $z_{11}=z_{21}=\ldots=z_{b1}$.
We consider the set $\mathcal{B}=\{z_{1,1}\}\cup \{z_{i,j}\}_{j \geq 2 ,\,\, 1\leq i \leq b}$.
For instance, when $m=2$ and $b=2$, we have 
$$\mathcal{B}=\{(e_1,e_1),(e_1,e_2),(e_1,e_3),(e_1,e_4),(e_2,e_1),(e_3,e_1),(e_4,e_1)\}.$$
Clearly, the cardinality of the set $\mathcal{B}$ is given by 
$$|\{ z_{i,j}\}_{1\leq b, \,\, 1 \leq j \leq 2^m }| - |\{z_{i,1}\}_{i\geq 2}|=2^mb-(b-1).$$

We claim that the set $\mathcal{B}$ is a basis for the subspace $\langle \mathrm{Im}(\varepsilon)\rangle$.\\
First, we prove that $\mathcal{B}$ is a linearly independent set. Suppose $z_{i,j}\in \mathcal{B}$ such that $(i,j)\neq (1,1)$. 
By definition of $\mathcal{B}$, the element $z_{i,j}$ is the unique element of $\mathcal{B}$ having a vector $e_j$ in position $i$.
 Thus, $z_{i,j}$ cannot be the linear combination (i.e. a sum) of any other vectors of $\mathcal{B}$ (see Lemma \ref{tecnico}).
Now, we have to consider the element $z_{1,1}$. Let $z_{1,1}=\sum_{(i,j)\in J} z_{i,j}$. W.l.o.g.,
we can assume by Lemma \ref{tecnico} that there is $(\bar{i},\bar{j})\in J$ such that $z_{\bar{i},\bar{j}}=(e_1,\ldots)$. 
Since $z_{\bar{i},\bar{j}}\not=z_{1,1}$ we can assume w.l.o.g. $z_{\bar{i},\bar{j}}=(e_1,e_{\bar{j}},e_1,\ldots,e_1)$, 
i.e. $\bar{i}=2$.
There is no other $z_{i,j}$ having $e_{\bar{j}}$ in the second position. 
Therefore, the sum $z_{1,1}$ should contain a $1$ in component $m+\bar{j}$, which is impossible.

Next, we prove that $\mathcal{B}$ generates $\langle \mathrm{Im}(\varepsilon)\rangle$. 
To do that, it suffices to prove that every element of $\mathrm{Im}(\varepsilon)$ 
belongs to the subspace generated by $\mathcal{B}$. 
If we consider an element $w=(e_{j_1},\ldots,e_{j_b})\in \mathrm{Im}(\varepsilon)$, we have
$$w=\left\{\begin{array}{ll} z_{1,j_1}+\cdots+z_{b,j_b} &  \textrm{if } b \textrm{ is odd,} \\ 
 z_{1,j_1} +\cdots+z_{b_,j_b}+z_{1,1}  & \textrm{if } b \textrm{ is even,}\end{array} \right.$$
 since
$$
	\begin{array}{c}
	 \, \\
	  \qquad \quad \, (e_{j_1},\,e_1,\ldots,\,e_1)+\\
	  b-1 \, \left\{
		\begin{array}{c}
					(\,e_1,e_{j_2},\ldots,\,e_1)+ \\
					\vdots \\
					(\,e_1,\ldots,\,e_1\,,e_{j_b})=
		\end{array} \right. \\
	  \hline 
		\quad \quad	\,(e_{j_1},e_{j_2},\ldots,e_{j_b})
	\end{array}
	b \, \,\textrm{odd}
	\qquad
	\begin{array}{c}
		\qquad \quad \, (\,e_1,\,e_1,\ldots,\,e_1)+ \\
	  \qquad \quad \, (e_{j_1},\,e_1,\ldots,\,e_1)+\\
		 b-1 \, \left\{
		\begin{array}{c}
					(\,e_1,e_{j_2},\ldots,\,e_1)+ \\
					\vdots \\
					(\,e_1,\ldots,\,e_1\,,e_{j_b})=
		\end{array} \right. \\
		\hline 
		\quad \quad	\,(e_{j_1},e_{j_2},\ldots,e_{j_b})
	\end{array}
	b \, \, \textrm{even}	
$$
\end{proof}
\end{proposition}

Let $\mathcal{A}$ be a subset of the plaintext set $\mathcal{M}$ such that 
$|\mathcal{A}|=\dim_{\FF_2}{(\langle \mathrm{Im}(\varepsilon)\rangle)}= 2^mb-(b-1)$. 
Let $a^i \in \mathcal{A}$, $1\leq i \leq|\mathcal{A}|$.
We construct the $(|\mathcal{A}|\times 2^mb)$-matrix $\mathbf{H}$ such that the 
$i$-th row is the image of the parallel map $\varepsilon$ applied to the plaintext $a^i \in \mathcal{A}$,
for $i \in \{1,\cdots, |\mathcal{A}|\}$:

\begin{equation}
\label{MatrixFirstRep}
\mathbf{H}=
\left( \begin{array}{c}
\varepsilon(a^1) \\
\varepsilon(a^2) \\
\vdots \\
\varepsilon(a^{|\mathcal{A}|}) \\
\end {array} \right)
=
\left( \begin{array}{cccc}
\varepsilon'(a_1^1) & \varepsilon'(a_2^1) & \cdots & \varepsilon'(a_b^1) \\
\varepsilon'(a_1^2) & \varepsilon'(a_2^2) & \cdots & \varepsilon'(a_b^2) \\
\vdots & \vdots & \vdots & \vdots \\
\varepsilon'(a_1^{|\mathcal{A}|}) & \varepsilon'(a_2^{|\mathcal{A}|}) & \cdots & \varepsilon'(a_b^{|\mathcal{A}|}) \\
\end {array} \right).
\end{equation}

We would like to determine the expected rank for such a matrix.
Generally speaking, for a random $(t \times n)$-matrix with entries in the finite field $\FF_q$, 
we can use the following well known results:
\begin{theorem}[\cite{CGC-alg-prep-migler}]
Let $t, k, n \in \NN \setminus \{0\}$, where $k \leq n$ and $k \leq t$.
\begin{enumerate}
\item The number of ordered $k$-tuples of linearly independent vectors in $(\FF_q)^n$ is 
      $$(q^n-1)(q^n-q)(q^n-q^2)\cdots(q^n-q^{k-1}).$$
\item The number of $k$-dimensional subspaces of $(\FF_q)^n$ is given by the $q$-binomial coefficient
    $${n \choose k}_q =\frac{\prod_{0 \leq i \leq k-1}(q^n-q^i)}{\prod_{0 \leq i \leq k-1}(q^k-q^i)}.$$ 
\item The number of $(t \times n)$-matrices of rank $k$ with entries in $\FF_q$ is given by the following formula
$$d_{k,t}={n \choose k}_q \, \prod_{0 \leq i \leq k-1}(q^t-q^i).$$
\end{enumerate} 
\end{theorem}
\noindent
We note that 
\begin{equation}
\label{coeffs}
\frac{{n \choose k-1}_q}{{n \choose k}_q}=\frac{q^k-1}{q^{n-k+1}-1}.
\end{equation}
By using the previous theorem, the relation in (\ref{coeffs}) and observing that 
$$ 
\frac{d_{t-2,t}}{d_{t,t}}=\frac{d_{t-2,t}}{d_{t-1,t}}\frac{d_{t-1,t}}{d_{t,t}}
$$
we immediately get the following corollary:
\begin{corollary}
Let $q=2$ and suppose $t <n$. We have the following relations:
\begin{eqnarray*}
d_{t,t} &=&(2^n-1)(2^n-2)\cdots(2^n-2^{t-1});\\
\frac{d_{t-1,t}}{d_{t,t}} &=& \frac{(2^t-1)}{(2^n-2^{t-1})}<\frac{1}{2^{n-t-1}}\leq 1;\\
\frac{d_{t-2,t}}{d_{t,t}} &=& \frac{(2^t-1)(2^{t-1}-1)}{3(2^n-2^{t-2})(2^{n}-2^{t-1})}.
\end{eqnarray*}
\end{corollary}
\begin{corollary}
\label{secondcor}
Let $q=2$ and suppose $t=n$. We have the following relations:
\begin{eqnarray*}
d_{n,n} &=&(2^n-1)(2^n-2)\cdots(2^n-2^{n-1});\\
\frac{d_{n-1,n}}{d_{n,n}} &=& \frac{2^n-1}{2^{n-1}}\approx 2 > 1;\\
\frac{d_{n-2,n}}{d_{n,n}} &=& \frac{(2^n-1)(2^{n-1}-1)}{9\cdot 2^{2n-3}}.
\end{eqnarray*}
\end{corollary}
In other words, the probability that a $(t \times n)$ random matrix $(t < n)$ with entries in $\FF_2$ has rank exactly
$t$ is significantly greater than the probability of having rank equal to $t-1$ or $t-2$ or less. 
Instead, the probability that a square $(n \times n)$ random matrix has rank $n-1$ is the greatest.

\begin{remark}
\label{CountingMatrixRank}
In theory, the previous theorem cannot be applied to our case because our construction imposes specific constraints, 
for example on the row-weight. 
However, in practice our ratio $\frac{d_{t-1,t}}{d_{t,t}}$ approaches that of the Corollary \ref{secondcor} for 
$t=\dim_{\FF_2}{(\langle \mathrm{Im}(\varepsilon)\rangle)}$.
\end{remark}

So, in order to point out the distribution of the ranks of our matrices we provide a bound on the number of the full-rank matrices.

\begin{lemma}
Let $c=2^m$, let $n=cb$ $(n \geq k)$ and $z=\dim_{\FF_2} \big(\langle \mathrm{Im}(\varepsilon) \rangle \big)$.
The total number of admissible vectors in $\langle \mathrm{Im}(\varepsilon) \rangle$ is $c^b$.
The average number $\xi(h)$ of admissible vectors in a subspace generated by $h$ linearly independent admissible vectors is
$$\xi(0)=0, \quad \xi(1)=1, \quad \xi(2)=2,$$
$$\xi(h)=h+(2^{h}-h-1)(\frac{c^b}{2^{z}}), \quad 3\leq h \leq z-1$$

\begin{proof}
An admissible vector can be any vector having weight $1$ in any of the $b$ components. 
There are $c^b$ such vectors.

The whole space $\langle \mathrm{Im}(\varepsilon) \rangle$ contains $2^z$ vectors.
The subspace ${\sf B}$ generated by $h$ independent vectors $(V_1, \ldots, V_h)$ contains $2^{h}$ vectors.
Of these, $h$ are $(V_1, \ldots, V_h)$ themselves (admissible) and one is the zero vector (non-admissible).
 
So ${\sf B}$ contains $2^{h}-h-1$ ``other'' vectors. 
To estimate how many of these are admissible, we simply multiply $2^{h}-h-1$ 
by the ratio $\frac{\textrm{admissible vectors}}{\textrm{all vectors}}=\frac{c^b}{2^{z}}$. 
Therefore, our average contains  $h+(2^{h}-h-1)\frac{c^b}{2^{z}}$ admissible vectors  
\end{proof}
\end{lemma}
\begin{theorem}
Let $c=2^m$, let $n=cb$ $(n \geq k \geq 1)$ and $z=\dim_{\FF_2} \big(\langle \mathrm{Im}(\varepsilon) \rangle \big)$.
\begin{enumerate}
\item The number of $(k\times n)$-matrices having rank $k$ can be estimated by the following formulas 
      $$\rho(k,k)=\prod_{1 \leq i \leq k}\left(c^b-\xi(i-1)\right) $$
      i.e.
      $$\rho(1,1)=c^b , \quad \rho(2,2)=c^b(c^b-1), \quad \rho(3,3)=c^b(c^b-1)(c^b-2),$$
      $$\rho(k,k)=c^b(c^b-1)(c^b-2)\prod_{4 \leq i \leq             
      k}\left(c^b-(i-1)-(2^{i-1}-i)\frac{c^b}{2^{z}}\right), 
      \quad k \geq 4 $$
\item The number of $(k\times n)$-matrices having rank $k-1$ can be estimated by the following recursive formula
      $$ \rho(2,1)=c^b,\quad \rho(3,2)=\rho(2,2)\xi(2)+\rho(2,1)(c^b-\xi(1))=3c^b(c^b-1),$$
      $$ \rho(4,3)=\rho(3,3)\xi(3)+\rho(3,2)(c^b-\xi(2))$$ 
      $$\rho(k,k-1)=\rho(k-1,k-1)\xi(k-1)+\rho(k-1,k-2)(2^z-2^{k-2})\frac{c^b}{2^z}, \quad k\geq 5,$$ 
      
\end{enumerate}

\begin{proof}
\begin{enumerate}
\item In order for a $(k \times n)$-matrix to have rank $k$, the rows must be linearly independent.
The first row can be any vector having weight $1$ in any of the $b$ component. There are $c^b$ such vectors, so $\rho(1,1)=c^b$
(i.e. $c^b$ is the total number of the admissible vectors).
The second row must be independent of the first row. That means it cannot be equal to the first row. 
There are $(c^b-1)$ choices for the second row and thus $\rho(2,2)=c^b(c^b-1)$.
The third row cannot be equal to one of the previous rows. 
But also, in our representation, it is impossible that two admissible vectors add to another admissible vector. 
Then we have $(c^b-2)$ choices for the third row, so $\rho(3,3)=c^b(c^b-1)(c^b-2)$.\\
On the other hand, if we add three or more admissible vectors we may get another admissible vector. 
As a consequence, if we are considering the $i$-th row, we must discard on average $\xi(i-1)$ vectors and so we can choose only among $c^b-\xi(i -1)$.
 
\item The set of the $(k \times n)$ matrices having rank exactly $k-1$ is the disjoint union of two sets:

\begin{itemize}
\item[a)] those having the first $k-1$ rows linearly independent (and so the $k$-th row dependent on the previous $k-1$ rows); 
\item[b)] those having the first $k-1$ rows linearly dependent  (and so these rows have rank $k-2$ and the $k$-th row is independent from them).
\end{itemize}
Therefore, the number of $(k \times n)$ matrices having rank exactly $k-1$ is obtained adding the following two values 
\begin{itemize}
\item[a)] the number of $(k-1)\times n$ matrices having rank $k-1$ multiplied by the number of all possible choices for the dependent row.\\
\item[b)] the number of $(k-1) \times n$ matrices having rank $k-2$ multiplied by the number of all possible choices for the independent row.
\end{itemize}

\begin{itemize}
\item The number of $(k-1)\times n$ matrices having rank $k-1$ is $\rho(k-1,k-1)$, for $k \geq 2$.
In case $k=2$, we have $\rho(1,1)=c^b$.
\item 
The number of all possible choices for the dependent row is $\xi(k-1)$ for $k \geq 2$; if $k=2$, the 
possible choice is exactly one, since the only second row we can choose is the first rows.
\item
The number of $(k-1) \times n$ matrices having rank $k-2$ is $\rho(k-1,k-2)$ and it makes sense for $k \geq 3$. When $k=2$ we have to consider a matrix having exactly one row and with rank $0$, so it is the {\em zero} row, but the {\em zero} row is not an admissible vector.
In other words, when we have only two rows, the set in $b)$ is empty.
In case $k=3$, we have $\rho(2,1)=c^b$, since the second row has to be equal to the first one.

\item The number of all possible choices for the independent row is \\
$(2^{z}-2^{k-2})(\frac{c^b}{2^z})$ and it is true for $k \geq 5$.
For $k=3$, we must choose a third row different from the first two. The first two are equal and so we have $c^b-1$ choices.
For $k=4$, we must choose a fourth row outside the space generated by the first three, but  
only two of the first three are distinct ans so we have $c^b-2$ choices.
\end{itemize}
Putting altogether we obtain our formula.

\end{enumerate}
\end{proof} 
\end{theorem}

\subsection{Application to AES}
\label{PrimaRepAES}
Because of the AES structure, we assign the following values to the parameters we have previously introduced. 
Let $V=(\FF_2)^{r}$ be our starting vector space with $r=128$ and $W=(\FF_2)^{s}$, $s>128$. We need to establish $s$.
We consider the quotient $\FF_{256} \cong \FF_2 [x]/\langle {\sf m} \rangle$,
where ${\sf m}=x^8+x^4+x^3+x+1 \in \FF_2[x]$ is the AES-polynomial. So $m=8$.  
According to the previous section, we consider $\varepsilon ': \FF_{2^8} \rightarrow (\FF_2)^{256}$ 
by means of a primitive element $\gamma$ of $\FF_{256}$, 
which is a root of the primitive polynomial\footnote{note that ${\sf n}\not={\sf m}$; 
we could not use ${\sf m}$ because it is not primitive.} ${\sf n}=x^8+x^4+x^3+x^2+1 \in \FF_2[x]$, and we define our parallel map 
$\varepsilon :V \rightarrow W$, with $r=m b=128$ and $s=2^m b=4096$, as
$$\varepsilon(v_1,\dots,v_{16})=(\varepsilon '(v_1),\dots, \varepsilon '(v_{16})).$$
\noindent
We have that $\dim_{\FF_2} \big(\langle \mathrm{Im}(\varepsilon)\rangle \big)= 4081$, by Proposition \ref{Dimension}.\\
A tipical round function of the AES cryptosystem consists of the composition of two parallel maps ({\sf AddRoundKey} 
and {\sf SubBytes} operations) and two non-parallel maps ({\sf ShiftRows} 
and {\sf MixColumns} operations).
We view the {\sf SubBytes} (and  {\sf AddRoundKey}) operation as a parallel map $\pi$  
\begin{eqnarray*}
 \pi : (\FF_{2^8})^{16} & \rightarrow & (\FF_{2^8})^{16}\\
       (y_1,\cdots,y_{16}) & \mapsto & (\pi_1(y_1),\cdots,\pi_{16}(y_{16}))
\end{eqnarray*}
where $y_i \in \FF_{2^8}$ and $\pi_{i} \in \Sym(\FF_{256})$, for $1\leq i \leq 16$.  
In the {\sf SubBytes} case, each component $\pi_i$, where $1 \leq i \leq 16$, 
is composition of inversion operation and an affine map; in the {\sf AddRoundKey} case, we have a sum with the round-key.
By the Theorem \ref{PermutPolyTheor} we recalled in the first section, 
we have that $\Sym(\FF_{256})= \langle ax+b,x^{254}\rangle$, where $a, b \in \FF_{256}$.
We note that a parallel map can be linearized using elementary results from Representation Theory.\\

Moreover, we claim that {\sf ShiftRows} is linear over $(\FF_{2})^{4096}$ and that {\sf MixColumns} is not linear over 
$(\FF_{2})^{4096}$, as follows. \\
First of all, we recall the map that describes the {\sf ShiftRows} operation:
\begin{small}
\begin{eqnarray*}
{\sf sh}: & (\FF_{2^8})^{16} & \rightarrow  (\FF_{2^8})^{16}\\
& (y_1,y_2,\cdots,y_{16})& \mapsto (y_1,y_6,y_{11},y_{16},y_5,y_{10},y_{15},y_4,y_{9},y_{14},y_3,y_{8},y_{13},y_{2},y_{7},y_{12}). 
\end{eqnarray*}
\end{small}
\noindent
Denoting by ${\bf y}=(y_1,\cdots,y_{16})$, we note that
$$\varepsilon({\bf y})=(\varepsilon'(y_1),\varepsilon'(y_2),\varepsilon'(y_3),\varepsilon'(y_4),\varepsilon'(y_5),\cdots,\varepsilon'(y_{16}))$$ 
and
$$\varepsilon ({\sf sh}({\bf y}))=
(\varepsilon'(y_1),\varepsilon'(y_6),\varepsilon'(y_{11}),\varepsilon'(y_{16}),\varepsilon'(y_5),\cdots ,\varepsilon'(y_{12})).$$
\noindent
The map ${\sf sh}$ is linearly extendible because $\sum_{i \in I}\varepsilon(b^i)=0$ clearly implies 
the following equality $\sum_{i \in I}\varepsilon({\sf sh}(b^i))=0$.

\pagebreak
\noindent
According to Lemma \ref{linear}, it is possible to construct the linear map 
$$A_{\mathrm{sh}}:(\FF_{2})^{4096}\rightarrow (\FF_{2})^{4096}$$
and so the {\sf ShiftRows} operation is linear over $(\FF_2)^{4096}$.\\ 

Now, we show that the {\sf MixColumns} operation is not linear over $(\FF_2)^{4096}$ using the following counterexample. 
\begin{example}
Let $w_1, w_2, w_3, w_4 \in W$ such that $w_1 + w_2+ w_3= w_4$:
\begin{eqnarray*}
w_1&=&(\varepsilon'(\gamma^1),\varepsilon'(\gamma^1),\varepsilon'(0),\varepsilon'(0),\varepsilon'(0),\cdots ,\varepsilon'(0))\\
w_2&=&(\varepsilon'(\gamma^1),\varepsilon'(0),\varepsilon'(\gamma^1),\varepsilon'(0),\varepsilon'(0),\cdots ,\varepsilon'(0))\\
w_3&=&(\varepsilon'(0),\varepsilon'(0),\varepsilon'(\gamma^1),\varepsilon'(0),\varepsilon'(0),\cdots ,\varepsilon'(0))\\
w_4&=&(\varepsilon'(0),\varepsilon'(\gamma^1),\varepsilon'(0),\varepsilon'(0),\varepsilon'(0),\cdots , \varepsilon'(0)).
\end{eqnarray*}
Now, we apply the {\sf MixColumns} operation ${\sf MC}$ to each vector $w_1, w_2, w_3, w_4$ obtaining the following
\begin{eqnarray*}
{\sf MC'}(w_1)&=&(\varepsilon'(\gamma^1),\varepsilon'(\gamma^3),\varepsilon'(0),\varepsilon'(\gamma^{51}),\varepsilon'(0),\cdots,\varepsilon'(0))\\
{\sf MC'}(w_2) &=&(\varepsilon'(\gamma^{3}),\varepsilon'(\gamma^{51}),\varepsilon'(\gamma^{3}),\varepsilon'(\gamma^{51}),\varepsilon'(0),\cdots,\varepsilon'(0))\\
{\sf MC'}(w_3)&=&
(\varepsilon'(\gamma^1),\varepsilon'(\gamma^3),\varepsilon'(\gamma^{51}),\varepsilon'(\gamma^1),\varepsilon'(0),\cdots,\varepsilon'(0))\\
{\sf MC'}(w_4)&=&(\varepsilon'(\gamma^3),\varepsilon'(\gamma^{51}),\varepsilon'(\gamma^1),\varepsilon'(\gamma^1),\varepsilon'(0),\cdots,\varepsilon'(0))
\end{eqnarray*}
where
$$
 \xymatrix{\ar @{} [dr] |{\circlearrowright}
   V \ar[d]^{{\sf MC}} \ar[r]^{\varepsilon} & W \ar[d]^{{\sf MC'}}\\
   V \ar[r]^{\varepsilon} & W}.
 $$
\ \\
Then we have that
${\sf MC'}(w_1)+{\sf MC'}(w_2)+{\sf MC'}(w_3)$ is
$$
(\varepsilon'(\gamma^3),
\varepsilon'(\gamma^{51}),
\varepsilon'(0)+\varepsilon'(\gamma^3)+\varepsilon'(\gamma^{51}),
\varepsilon'(\gamma^1),
\varepsilon'(0),\cdots,\varepsilon'(0)).$$

The third component of the previous vector is a sum in $(\FF_2)^{256}$ and it has weight equal to $3$.
So, the vector ${\sf MC'}(w_1)+{\sf MC'}(w_2)+{\sf MC'}(w_3)$ is an element of the admissible space but it is a non-admissible 
vector.\\
Therefore,  
$ {\sf MC'}(w_4)={\sf MC'}(w_1+w_2+w_3)\not={\sf MC'}(w_1)+{\sf MC'}(w_2)+{\sf MC'}(w_3)$
and so the {\sf MixColumns} is not linear over $W$.
It means that the extension of ${\sf MC}$ is not linearly extendible.
\end{example}

\begin{remark}
If all the AES operations were parallel maps,
it would be possible to linearize the ``full'' cryptosystem because the set of the parallel maps is a group with respect to 
the composition operation.
\end{remark}

\subsection{Application to PRESENT}
As for AES, we assign the right values to our parameters, according to PRESENT's structure.
Let $V=(\FF_2)^{r}$ be our starting vector space with $r=64$, and $W=(\FF_2)^{s}$ with $s>64$.
We consider $\varepsilon ': \FF_{2^4} \rightarrow (\FF_2)^{16}$ and we define our parallel map 
$\varepsilon :V \rightarrow W$, with $r=m b=64$ and $s=2^m b=256$, as
$$\varepsilon(v_1,\dots,v_{16})=(\varepsilon '(v_1),\dots, \varepsilon '(v_{16})).$$
We note that $\dim_{\FF_2} \big(\langle \mathrm{
Im}(\varepsilon) \rangle \big)= 241$ (see Proposition \ref{Dimension}).\\
A typical round function of the PRESENT cryptosystem consists of the composition of two parallel maps ({\sf addRoundKey} 
and {\sf sBoxLayer} operations) and one non-parallel map ({\sf pLayer} operation).
The {\sf addRoundKey} (and {\sf sBoxLayer}) operation is a parallel maps $\pi$  
\begin{eqnarray*}
 \pi : (\FF_{2^4})^{16} & \rightarrow & (\FF_{2^4})^{16}\\
       (t_1,\cdots,t_{16}) & \mapsto & (\pi_1(t_1),\cdots,\pi_{16}(t_{16}))
\end{eqnarray*}
where $\pi_{i} \in \Sym(\FF_{16})$. 
In the {\sf sBoxLayer} case, each component $\pi_i$ ($1 \leq i \leq 16$) is given by the table in Subsection \ref{present description}; 
when $\pi$ is the {\sf addRoundKey} operation, we have only a bitwise sum with the round-key.\\
Moreover, it is easy to see that  {\sf pLayer} is not linear over $(\FF_{2})^{256}$. 
\begin{example}
Let $w_1, w_2, w_3, w_4 \in W$ such that $w_1 + w_2+ w_3= w_4$ and let $\zeta,\eta, \vartheta,\xi, \mu$ be distinct non-zero 
elements in $\FF_{2^4}$. Suppose that 
\begin{eqnarray*}
w_1&=&(\varepsilon'(\zeta),\varepsilon'(\zeta),\varepsilon'(0),\varepsilon'(0),\varepsilon'(0),\cdots ,\varepsilon'(0))\\
w_2&=&(\varepsilon'(\zeta),\varepsilon'(0),\varepsilon'(\zeta),\varepsilon'(0),\varepsilon'(0),\cdots ,\varepsilon'(0))\\
w_3&=&(\varepsilon'(0),\varepsilon'(0),\varepsilon'(\zeta),\varepsilon'(0),\varepsilon'(0),\cdots ,\varepsilon'(0))\\
w_4&=&(\varepsilon'(0),\varepsilon'(\zeta),\varepsilon'(0),\varepsilon'(0),\varepsilon'(0),\cdots , \varepsilon'(0)).
\end{eqnarray*}
Now, we apply the {\sf pLayer} transformation ${\sf pL}$ to each vector $w_1, w_2, w_3, w_4$ obtaining the following

\begin{eqnarray*}
{\sf pL'}(w_1)&=&(\varepsilon'(\eta),{\bf \varepsilon'(0)}_3,\varepsilon'(\eta),{\bf \varepsilon'(0)}_3,
                \varepsilon'(\eta),{\bf \varepsilon'(0)}_3,\varepsilon'(\eta),{\bf \varepsilon'(0)}_3)\\
{\sf pL'}(w_2)&=&(\varepsilon'(\vartheta),{\bf \varepsilon'(0)}_3,\varepsilon'(\vartheta),{\bf \varepsilon'(0)}_3, 
                \varepsilon'(\vartheta),{\bf \varepsilon'(0)}_3,
                  \varepsilon'(\vartheta),{\bf \varepsilon'(0)}_3)\\
{\sf pL'}(w_3)&=&(\varepsilon'(\xi),{\bf \varepsilon'(0)}_3,\varepsilon'(\xi),{\bf \varepsilon'(0)}_3,
                 \varepsilon'(\xi),{\bf \varepsilon'(0)}_3,\varepsilon'(\xi),{\bf \varepsilon'(0)}_3)\\
{\sf pL'}(w_4)&=&(\varepsilon'(\mu),{\bf \varepsilon'(0)}_3,\varepsilon'(\mu),{\bf \varepsilon'(0)}_3,
                 \varepsilon'(\mu),{\bf \varepsilon'(0)}_3,\varepsilon'(\mu),{\bf \varepsilon'(0)}_3)
\end{eqnarray*}

where ${\bf \varepsilon'(0)}_3$ means $(\varepsilon'(0),\varepsilon'(0),\varepsilon'(0))$.
Then, we have that
\begin{small}
$${\sf pL'}(w_4)={\sf pL'}(w_1+w_2+w_3)\not= 
{\sf pL'}(w_1)+{\sf pL'}(w_2)+{\sf pL'}(w_3)=(\varepsilon'(\eta)+\varepsilon'(\vartheta)+\varepsilon'(\xi),\ldots),$$
\end{small}
where the first component has weight $3$, and so the {\sf pLayer} is not a linear operation over $W$.
\end{example}

\begin{remark}
As in the AES case, if all the PRESENT's operations were parallel maps, 
it would be possible to linearize the ``full'' cryptosystem because the set of the parallel maps is a group with respect 
to the composition operation.
\end{remark}

\subsection{Application to SERPENT}
Let $V=(\FF_2)^r$ be our starting vector space with $r=128$. In order to identify the value of $r \geq s$, where $W=(\FF_2)^s$, 
we have to consider the map $$\varepsilon':(\FF_{2^4})\rightarrow (\FF_{2})^{2^4} .$$
We define our parallel map $\varepsilon: V \rightarrow W$ with $r=m b=128$ and $s=2^m b=512$ as 
$$
\varepsilon(v_1, \ldots,v_{32})=(\varepsilon'(v_1), \ldots, \varepsilon'(v_{32})).
$$
Note that $\dim_{\FF_2}(\langle \mathrm{Im}(\varepsilon)\rangle)=2^mb-(b-1)=481$.\\
The components of a typical round function are the parallel $S$-box, the affine transformation described in Subsection \ref{serpent-description} 
and the translation with the round key. Obviously, key translation and $S$-box are parallel maps of type
\begin{eqnarray*}
\pi:(\FF_{2^4})^{32} &\rightarrow &(\FF_{2^4})^{32}\\
(t_1,\ldots,t_{32})  & \mapsto    &(\pi_1(t_1),\ldots,\pi_{32}(t_{32}))
\end{eqnarray*}
where $\pi_i \in \Sym(\FF_{2^4})$.
Similarly to what was done for AES and PRESENT, we could provide a counterexample to show that the linear transformation of 
SERPENT is not linear over $(\FF_2)^{512}$.

\section{Results on a larger embedding}
\label{OrbitRepresentation}

In this section we provide another specific embedding that can be seen as an improvement of the former (\ref{defepsilon}).
Also the new embedding can be applied to AES, PRESENT and SERPENT. 
In Subsection \ref{FirstRepresent} we considered $\Omega=V$ as a vector space 
and we found an embedding $V \hookrightarrow W$ such that the $S$-boxes and the key-additions become linear. 
However, in this way we lost the linearity of the Mixing Layer $\lambda$ and so here 
we make a larger embedding where the linearity of $\lambda$ is recovered, without losing the linearity of the key addition. 
We do lose the linearity of the $S$-boxes, but their non-linearity is probably kept low.     

Starting from the setting  we described in the previous section, we consider our parallel map 
$\varepsilon :(\FF_{2^m})^b \rightarrow ((\FF_2)^{2^m})^{b}$  defined as
$\varepsilon(v_1,\dots,v_{b})=(\varepsilon '(v_1),\dots, \varepsilon '(v_{b})).$

\pagebreak
Now, let ${\bf M}$ be a matrix in $\GL((\FF_{2})^{mb})$ and let $t$ be its order, ${\bf M}^t=\mathrm{id}_V$.
Let $V=(\FF_2)^r$ be a vector space with dimension $r=m b$ and let $W=(\FF_2)^s$ be the vector space with dimension $s=2^m b t$.
The space embedding $\alpha:V \rightarrow W$ we propose in this section is defined as follows
\begin{equation}
\label{orbita}
\alpha(v)=(\varepsilon(v),\varepsilon ({\bf M} v),\dots, \varepsilon({\bf M}^{t-1} v)).
\end{equation}
From now on, $\alpha$ denotes the map in (\ref{orbita}). 
Thanks to Proposition \ref{Dimension}, we can easily prove the following proposition:
\begin{proposition}
\label{DimOrbitaBound}
Let $V=(\FF_2)^r$ be a vector space with dimension $r=m b$ and let $W=(\FF_2)^s$ be the vector space with dimension $s=2^m b t$.
Let $\alpha$ be as in (\ref{orbita}).
Then we have $$2^m b-(b-1)\leq \dim_{\FF_2}\big(\langle \mathrm{Im}(\alpha) \rangle \big)\leq (2^m b-(b-1))t$$
\begin{proof}
By Proposition \ref{Dimension}, $\dim_{\FF_2} \big(\langle \mathrm{Im}(\varepsilon) \rangle \big)= 2^m b -(b-1)$. Since
$$\{(\varepsilon(v),\varepsilon ({\bf M} v),\dots, \varepsilon({\bf M}^{t-1} v))\mid v \in V \}\subset \{(\varepsilon(v_1),\dots, \varepsilon(v_t))\mid v_1, \ldots, v_t \in V\},$$
then $$\dim_{\FF_2}\big(\langle \mathrm{Im}(\alpha) \rangle \big)\leq (2^m b-(b-1))t.$$
On the other hand, considering the projection of $\{(\varepsilon(v),\varepsilon ({\bf M} v),\dots, \varepsilon({\bf M}^{t-1} v))\}$ 
on the first component (the first $b$ bytes), the lower bound follows immediately, again considering Proposition \ref{Dimension}.
\end{proof}
\end{proposition}
We can further improve Proposition \ref{DimOrbitaBound} for byte-oriented Mixing Layer. 
\begin{proposition}
\label{DimensionOrbita} 
Let $V=(\FF_2)^r$ be a vector space with dimension $r=m b$ and let $W=(\FF_2)^s$ be the vector space with dimension $s=2^m b t$. 
Let ${\bf M}\in \GL((\FF_{2^m})^{b})$.
Let $\alpha$ be as in (\ref{orbita}).
Then we have $$\dim_{\FF_2}\big(\langle \mathrm{Im}(\alpha)\big)\leq 2^{m}bt -(bt-1)- mb(t-1)$$
\begin{proof}
Let $T=\langle \mathrm{Im}(\alpha)\rangle$. 
For any $w_1,w_2 \in W$, let $w_1 \cdot w_2$ denote their scalar product. 
It is sufficient to show that there exist $(bt-1)+ mb(t-1)$ elements in $T^{\bot}$ that are linearly independent, 
where $T^{\bot}=\{w \in W \mid w \cdot {\bf t}=0, \forall {\bf t}\in T\}$ is the orthogonal space of $T$ 
(or the ``dual'' of $T$, in coding theory notation). In fact, this means 
$$
\dim T^{\bot} \geq (bt-1)+ mb(t-1)
$$ 
and since $\dim T= \dim W -\dim T^{\bot},$ our result could follows.

Consider the following matrix product with ${\bf M}=(a_{i,j})$
\begin{displaymath}
\left( \begin{array}{ccccc}
a_{11} & a_{12} & \cdots & \cdots & a_{1b} \\
a_{21} & a_{22} & \cdots & \cdots & a_{2b} \\
\vdots & \vdots & \vdots & \vdots & \vdots \\
a_{b1} & a_{b2} & \cdots & \cdots & a_{bb} \\  
\end{array} \right)
\left( \begin{array}{c}
v_1\\
v_2\\
\vdots\\
v_b\\
\end{array} \right)
=
\left( \begin{array}{c}
v'_{1}\\
v'_{2}\\
\vdots\\
v'_{b}\\
\end{array} \right)
\end{displaymath}
\noindent
Obviously, $v'_{1}=\sum_{i=1}^{b}v_{i}a_{1i}$.\\
Let $S'$ be a subspace of $(\FF_2)^m$ such that $\dim(S')=m-1$. 
For any $1 \leq i\leq b$, let $S_i=\{\beta \in (\FF_{2^m})\mid \beta a_{1i} \in S'\}$.
We note that $S_i$ is a subspace and that 
$$\Big\{\sum_{i=1}^{b}v_{i}a_{1i} \mid v_{i} \in S_i,\, 1 \leq i\leq b \Big\}=S'$$ 
and that $|S_i|=|S'|=2^{m-1}$.
There exists a bijection via orthogonality between the
sets $\mathcal{S}=\{S < (\FF_2)^m | \dim(S)=m-1\}$ 
and $\{S^{\bot} < (\FF_2)^m | \dim(S^{\bot})=1\}$; their cardinality is obviously $2^m-1$.
We can choose a linear basis for $\mathcal{S}\cup\{0\}$, i.e. 
$\mathcal{S}\cup\{0\}=\langle{\bf e}_1^{\bot},\ldots,{\bf e}_m^{\bot}\rangle$. 
Therefore, each row of ${\bf M}$ generates $m$ linearly independent elements of $T^{\bot}$. \\
Two relations coming from two different rows are independent, since the matrix ${\bf M}$ has full rank, 
for a total of $mb$ relations.

Now, we construct the elements of the orthogonal space that correspond to the relations induced by the rows of ${\bf M}$.
We are considering the case $(v, {\bf M}v)$ and we observe that 
\begin{equation}
\label{voneprime}
\sum_{i=1}^{b}v_{i}a_{1i}=v'_{1}=({\bf M}v)_1
\end{equation}
where $v_i \in S_i$. 
Since $\varepsilon'(S_i)\subset(\FF_2)^{2^m}$, we consider 
$w_i=\sum_{\ell \in \varepsilon'(S_i)} \ell$ where  $\mathrm{w}(w_i)=|\varepsilon'(S_i)|=2^{m-1}$ and $w_i \in (\FF_2)^{2^m}$.
The element of $T^{\bot}$ coming from (\ref{voneprime}) and $S$ is 
$$
(w_1,\ldots, w_b, w'_1,\ldots,0,\ldots 0)
$$
where $w'_1=\sum_{\ell \in \varepsilon'(S')} \ell$. Clearly, $m-1$ similar elements come from (\ref{voneprime}) and $\mathcal{S}$.\\
If we consider the relations given by the $h$-th row of ${\bf M}$, i.e. $\sum_{i=1}^{b}v_{i}a_{hi}=v'_{h}$, 
we obtain the following elements
$$
(w_1,\ldots, w_b, 0,\ldots, w'_h,\ldots,0,\ldots 0).
$$
At this point, we have constructed the $mb$ elements of the orthogonal space corresponding to the previous relations.\\
Instead of considering $(v, {\bf M}v)$, since clearly ${\bf M}({\bf M}^i v)={\bf M}^{i+1}v$, we 
can apply the previous construction to each pair $({\bf M}^iv, {\bf M}^{i+1}v)$, for $1\leq i \leq t-2$, 
obtaining the corresponding elements
\begin{equation}
\label{firstelement}
(\underbrace{0,\ldots, 0}_{b(i-1)},\underbrace{w_1,\ldots,w_b}_{b}, \underbrace{0,\ldots,w'_h,\ldots, 0}_{b},\underbrace{0,\ldots, 0}_{bt-(i+1)b})
\end{equation}
We have found exactly $mb(t-1)$ vectors in $T^{\bot}$. 
Since the pairs $({\bf M}^{i}v, {\bf M}^{i+1}v)$ and $({\bf M}^{j}v, {\bf M}^{j+1}v)$ with $i \not=j$ involve different bytes, 
the relations given by $({\bf M}^{i}v, {\bf M}^{i+1}v)$ are independent from those given by 
$({\bf M}^{j}v, {\bf M}^{j+1}v)$.
Then we have $mb(t-1)$ independent relations (i.e. linearly independent elements of the orthogonal space). 

Thanks to Proposition \ref{Dimension}, we have exactly $(bt-1)$ further relations, corresponding to elements in $T^{\bot}$ of type
\begin{equation}
\label{secondelement}
(\underbrace{0,\ldots, 0}_{k-1},\underbrace{1,\ldots,1}_{b},\underbrace{1,\ldots,1}_{b},\underbrace{0,\ldots ,0}_{bt-(k+1)})
\end{equation}
with $1\leq k \leq bt$. 
  
The vectors (\ref{firstelement}) and (\ref{secondelement}) form clearly a linearly independent set. 
\end{proof}
\end{proposition}

As we have done in previous section, we can construct the following matrix.
Let $\mathcal{D}$ be a subset of the plaintext set $\mathcal{M}$ such that 
$|\mathcal{D}|=\dim_{\FF_2}{(\langle \mathrm{Im}(\alpha)\rangle)}$. 
Let $a_i \in \mathcal{D}$, $1\leq i \leq|\mathcal{D}|$. 
We construct the $(|\mathcal{D}|\times 2^mbt)$-matrix $\mathbf{D}$ such that the $i$-th row is the image of the map 
$\alpha$ applied to the plaintext $a_i \in \mathcal{D}$, for $i \in \{1,\cdots, |\mathcal{D}|\}$:
\begin{displaymath}
\mathbf{D}=
\left( \begin{array}{c}
\alpha(a^1) \\
\alpha(a^2) \\
\vdots \\
\alpha(a^{|\mathcal{D}|}) \\
\end {array} \right)
=
\left( \begin{array}{cccc}
\varepsilon(a^1) & \varepsilon({\bf M}a^1) & \cdots & \varepsilon({\bf M}^{t-1}a^1) \\
\varepsilon(a^2) & \varepsilon({\bf M}a^2) & \cdots & \varepsilon({\bf M}^{t-1}a^2) \\
\vdots & \vdots & \vdots & \vdots \\
\varepsilon(a^{|\mathcal{D}|}) & \varepsilon({\bf M}a^{|\mathcal{D}|}) & \cdots & \varepsilon(M^{t-1}a^{|\mathcal{D}|}) \\
\end {array} \right).
\end{displaymath}
\begin{remark}
We expect the rank of this matrix to have a behavior similar to that of matrix $\mathbf{H}$ (\ref{MatrixFirstRep}), 
see Remark \ref{CountingMatrixRank}. Our experiments confirm this.
\end{remark}

Let $\tilde{\mathcal{G}}$ be the set of parallel maps $\tilde{\pi}:(\FF_{2^m})^b \rightarrow (\FF_{2^m})^b$, such that, 
for any $1\leq j \leq b$,
$\tilde{\pi}_j(x)=ax+c$, with $a\not=0, c \in \FF_{2^m}$ ($a$ and $c$ do not depend on $j$).\\
Let $\bar{\mathcal{G}}$ be the set of parallel maps $\bar{\pi}:(\FF_{2^m})^b \rightarrow (\FF_{2^m})^b$, such that, 
for any $1\leq j \leq b$,
$\bar{\pi}_j(x)=x+d_j$, with $d_j \in \FF_{2^m}$.\\
Note that both  $\tilde{\mathcal{G}}$ and  $\bar{\mathcal{G}}$ are subgroups of $\Sym((\FF_{2^m})^b)$ and we define $\mathcal{G}$ as  
$$ \mathcal{G}=\left\langle \tilde{\mathcal{G}},\bar{\mathcal{G}},{\bf M}\right\rangle < \Sym((\FF_{2^m})^b).$$

The following result holds:

\begin{proposition}
\label{linearG}
Let $\sigma$ be  either an element of $\tilde{\mathcal{G}}$ or an element of $\bar{\mathcal{G}}$, 
then there exists $A_{\sigma}:W \rightarrow W$ which is linear.
 \begin{proof}
 We want to apply Lemma \ref{linear} and so we must only show that $\sigma $ is li\-near\-ly extendible.
 Let $\{v^i\}_{i \in I} \subset V$ such that $\sum_{i \in I}{\alpha(v^i)}=0$, we have to prove that
 $\sum_{i \in I}{\alpha(\sigma(v^i))}=0$. 
Note that $\sum_{I}{\alpha(v^i)}=0$ is equivalent to
 {\small
 \begin{equation*}
\sum_{I}{\left(\varepsilon'(v^i)_{1},\dots,\varepsilon'(v^i)_{b},\varepsilon' ({\bf M} v^i)_{1},\dots,\varepsilon' ({\bf M} v^i)_{b},\dots,       
 \varepsilon'({\bf M}^{t-1} v^i)_{1},\dots,\varepsilon'({\bf M}^{t-1} v^i)_{b}\right)}=0.
 \end{equation*}
 }
 Then we have the following system $S_j$ for any $1\leq j\leq b$
 $$
 S_j=
 \begin{cases}
  \sum_{I}{\varepsilon'(v^{i}_j)}=0\\
  \sum_{I}{\varepsilon'(({\bf M}v^{i})_j)}=0\\
  \vdots \\
  \sum_{I}{\varepsilon'(({\bf M}^{t-1}v^{i})_j)}=0 .
 \end{cases}
 $$
 Using Lemma \ref{parity}, we have that $S_j$ is equivalent to  $S_j'$
 $$
 S_j'=
 \begin{cases}
  |\{\ell \mid v^{\ell}_{j}=v^{i}_{j}\}| \,\, \text{is even} \,\, \forall i \in I\\
  |\{\ell \mid ({\bf M}v^{\ell})_{j}=({\bf M}v^{i})_{j}\}|\,\, \text{is even} \,\, \forall i \in I\\
  \vdots \\
  |\{\ell \mid ({\bf M}^{t-1}v^{\ell})_{j}=({\bf M}^{t-1}v^{i})_{j}\}| \,\,\text{is even} \,\, \forall i \in I. 
 \end{cases}
 $$
 Suppose $\sigma \in \tilde{\mathcal{G}}$ which means that $\sigma(v)=\sigma(v_1,\cdots,v_b)=(\sigma_1(v_1),\cdots,\sigma_b(v_b))$ 
where
 $\sigma_i(v_i)=av_i+c$ for any $1\leq i \leq b$ and $a\not=0, c \in \FF_{2^m}$.\\		 
 Since ${\bf M}$ is linear, we have
 \begin{eqnarray*}
   ({\bf M}^h\sigma(v^{\ell}))_j & = & ({\bf M}^h(av_1^{\ell}+c,\cdots,av_b^{\ell}+c))_j\\
                                 & = & (a{\bf M}^hv^{\ell}+{\bf M}^h(c,\cdots,c))_j\\
                                 & = & (a{\bf M}^hv^{\ell})_j+({\bf M}^h(c,\cdots,c))_j\\
                                 & = & a({\bf M}^hv^{\ell})_j +\bar{c},
 \end{eqnarray*}
 where $\bar{c}$ is a constant {\em independent} of $\ell$.
 \noindent
 We have that, $\forall i \in I$ and for any $1 \leq h \leq t-1$, $|\{\ell \mid ({\bf M}^{h}v^{\ell})_{j}=({\bf M}^{h}v^{i})_{j}\}| $ is even  
 and so that
 $|\{\ell \mid a({\bf M}^{h}v^{\ell})_{j}+\bar{c}=a({\bf M}^{h}v^{i})_{j}+\bar{c}\}| \,\,\text{is even}$.
 Thanks to Lemma \ref{parity}, our thesis follows.
 
 Suppose now that $\sigma \in \bar{\mathcal{G}}$, i.e. $\sigma(v)=v+d$ for some $d \in V$.
 Since 
 \begin{eqnarray*}
   ({\bf M}^h\sigma(v^{\ell}))_j & = & ({\bf M}^h(av^{\ell}+d))_j\\
                                 & = & ({\bf M}^hv^{\ell})_j+({\bf M}^h(d))_j\\
                                 & = & ({\bf M}^hv^{\ell})_j +\bar{d},
 \end{eqnarray*}
 where $\bar{d}$ is a constant {\em independent} of $\ell$
 and $|\{\ell \mid ({\bf M}^{h}v^{\ell})_{j}=({\bf M}^{h}v^{i})_{j}\}|$ is even, we have that
 $$|\{\ell \mid ({\bf M}^{h}v^{\ell})_{j}+\bar{d}=({\bf M}^{h}v^{i})_{j}+\bar{d}\}|$$ is even.
 By Lemma \ref{parity}, our thesis follows.
\end{proof}
\end{proposition}


\subsection{Application to AES}
\label{AESOrbit}
Let $V=(\FF_2)^r$ be a vector space with dimension $r=128$ and let 
${\sf M}:V \rightarrow V$ be the {\sf MixingLayer} of AES, that is, the composition of {\sf ShiftRows}  and {\sf MixColumns}. 
Since ${\sf M}$ has order equal to $8$ (i.e. ${\sf M}^8=\mathrm{id}_V$), the map $\alpha:V \rightarrow W$ we propose is defined 
as follows
\begin{equation}
\label{AESorbita}
\alpha(v)=(\varepsilon(v),\varepsilon ({\sf M} v),\dots, \varepsilon({\sf M}^7 v)),
\end{equation}
where $W=(\FF_2)^s$ is the vector space with dimension $s=2^m b t=2^{15}$ and $\varepsilon$ is the map defined in 
Subsection \ref{PrimaRepAES}: $\varepsilon :(\FF_2)^{128}\rightarrow (\FF_2)^{4096}$.\\
Let $T=\langle\mathrm{Im}(\alpha)\rangle$ with $\alpha$ in (\ref{AESorbita}). We can easily determine $\dim(T)$.
\begin{fact}
\label{AESdimensionOrbit}
In the AES case we have  $$\dim_{\FF_2}(T)= 2^{m}bt -(bt-1)- mb(t-1)=31745.$$
\begin{proof}
Let $\lambda = 2^mbt-(bt-1)-mb(t-1)$. By computational experiments, we have found a $(\lambda \times 2^mbt)$ 
full rank matrix for the $\alpha$ representation in the AES case. Which means $\dim_{\FF_2} T \geq \lambda$. 
Thanks to Proposition \ref{DimensionOrbita} we conclude that $\dim_{\FF_2} T = \lambda.$ 
\end{proof}
\end{fact}

We note that the group 
$$ \mathcal{G}=\left\langle \tilde{\mathcal{G}},\bar{\mathcal{G}},{\sf M}\right\rangle < \Sym((\FF_{2^8})^{16}).$$ 
contains all the permutations of the AES-round function, except notably for the $S$-box operation.
  
\begin{proposition}
Let ${\sf M}$ be the {\sf MixingLayer}. Then $\alpha$ is a space embedding with respect to 
$\mathcal{G}=\left\langle \tilde{\mathcal{G}},\bar{\mathcal{G}},{\sf M}\right\rangle$.
\begin{proof}
According to Proposition \ref{linearG}, there exists a linear map $A_{\sigma}: W \rightarrow W$ in case $\sigma$ is  
$\tilde{\mathcal{G}}$ or $\bar{\mathcal{G}}$. We note that the previous result is independent from ${\bf M}$. 
Let ${\bf M}$ be the {\sf MixingLayer} ${\sf M}$. 
Since $\alpha(v^i)=(\varepsilon(v^i),\varepsilon ({\sf M} v^i),\dots, \varepsilon({\sf M}^{7} v^i))$
and
\begin{eqnarray*} 
\alpha({\sf M}v^i)& =& (\varepsilon({\sf M}v^i),\varepsilon ({\sf M}^2 v^i),\dots, \varepsilon({\sf M}^{8} v^i))\\
                   &=&  (\varepsilon({\sf M}v^i),\varepsilon ({\sf M}^2 v^i),\dots, \varepsilon(v^i))
\end{eqnarray*}
$\alpha({\sf M}v^i)$ is a permutation of $\alpha(v^i)$. Obviously, we have that $\sum_{i \in I}\alpha(v^i)=0$ implies 
$\sum_{i \in I}\alpha({\sf M}v^i)=0$.
\end{proof}
\end{proposition}

With a fixed $K$, the encryption $\phi_K$ is the composition of {\sf AddRoundKey}, {\sf Subbytes}  and {\sf MixingLayer}. 
So the only part of $\phi_K$ which is not linear (with our map $\alpha$) is the {\sf SubBytes} operation.

\subsection{Application to PRESENT}
Let $V=(\FF_2)^r$ be a vector space with dimension $r=64$ and let 
${\sf M}:V \rightarrow V$ be the {\sf pLayer} of PRESENT. 
Since ${\sf M}^3=\mathrm{id}_V$, the map $\alpha:V \rightarrow W$ we propose is defined as follows
\begin{equation}
\label{PRESENTorbita}
\alpha(v)=(\varepsilon(v),\varepsilon ({\sf M} v),\varepsilon({\sf M}^2 v)),
\end{equation}
where $W=(\FF_2)^s$ is the vector space with dimension $s=2^m b t=768$.
Let $\alpha$ be as in (\ref{PRESENTorbita}) and $T=\langle \mathrm{Im}(\alpha)\rangle$.
Also in this case it is possible to prove (with a computation) that $\dim_{\FF_2}(T)= 2^{m}bt -(bt-1)- mb(t-1)=593$

With a fixed $K$, the encryption $\phi_K$ is the composition of {\sf addRoundKey}, {\sf sBoxLayer} and {\sf pLayer}. 
So the only part of $\phi_K$ which is not linear (with our map $\alpha$) is the {\sf sBoxlayer} operation.

\subsection{Application to SERPENT}
Let $V=(\FF_2)^r$ be a vector space with dimension $r=128$ and let 
${\sf M}:V \rightarrow V$ be the affine transformation of SERPENT. 
Since the order of ${\sf M}$ is 
greater\footnote{to be precise it is $110329570561973845861261474090270635$, as computed directly with MAGMA.} 
than $2^{116}$, it is huge and impractical to consider the map $\alpha:V \rightarrow W$ 
\begin{equation}
\alpha(v)=(\varepsilon(v),\varepsilon ({\sf M} v),\dots, \varepsilon({\sf M}^{2^{80}} v),\dots).
\end{equation}
since $W=(\FF_2)^s$ would have $s=2^m b t>2^4 \cdot 32 \cdot 2^{116}=2^{125}$, making the rank computation 
impossible with nowadays technology. 


\section{Attack strategies} 
\label{strategies}
In this paper we do not report on successful attacks on (full
versions of) the AES or other well-known ciphers.
It is true that we have implemented several attacks aiming at distinguishing AES from random permutations, presented in some
talks, and that we have collected
some data indicating that our approach is likely to succeed.
Yet, our data do not provide an overwhelming statistical evidence
for the full cipher versions.
Therefore, in this section we sketch some attack strategies that we have
followed, without giving full details.

The most difficult task in assessing the success of one
of our embeddings is, by far, to estimate the non-linearity decrease
of the cryptosystem. For example, a rigorous determination of the $s$-extendibility
(Subsection \ref{Weakernotion}) appears completely out of reach.
The only methods we can use to estimate the non-linearity fall are
"a posteriori" checks on linear dependences.

We have implemented only chosen-plaintex attacks, either with single-key
or with related keys.
In the single-key scenario,  we proceed in three steps:

\begin{enumerate}
\item we choose a set $S$ of $N$ ($31745 \times 2^{15}$)-matrices, with rows taken from $T$ 
(Fact \ref{AESdimensionOrbit});
\item we encrypt all matrices in $S$ (row by row) with a given key and compute their ranks;
\item we compare their rank distribution with the expected rank distribution
      for a set of $N$ {\bf random} ($31745 \times 2^{15}$)-matrices, with rows taken from $T$, 
      aiming at distinguishing the two distributions;
\item to validate the distinguishing statistical test, we also create sets of 
$N$ {\bf random} ($31745 \times 2^{15}$)-matrices (in $T$) and we compare them with the expected distribution, aiming at {\em not} distinguishing them.      
\end{enumerate}

In the related-key scenario we proceed similarly.
Let $n_k$ be the number of related keys:

\begin{enumerate}
\item we choose a set $S$ of $N$ ($31745 \times 2^{15}$)-matrices, with rows taken from $T$;
\item we encrypt all matrices in $S$ (row by row) with all keys and compute their ranks;
\item we compare their rank distribution with the expected rank distribution
      for a set of $N N_k$ {\bf random} ($31745 \times 2^{15}$)-matrices, with rows taken from $T$, 
      aiming at distinguishing the two distributions;
\item to validate the distinguishing statistical test, we also create sets of 
$N n_k$ {\bf random} ($31745 \times 2^{15}$)-matrices (in $T$) and we compare them with the expected distribution, aiming at {\em not} distinguishing them.      
\end{enumerate}

Apart from the obvious difference in the dealing of 
the single-key/related-key mechanism, the two scenarios are very similar,
since in both we hope to spot a significant deviation by looking at ranks.
Matrix ranks do depend on the linear dependences of the rows and are
much easier to compute and compare, so they are cheap indicators
for the non-linearity behavior (see Marsaglia's test, e.g. \cite{CGC-cry-art-SotoRTA},\cite{CGC-misc-NIST-SP80022}).\\
On the other hands, since a great deal of row dependences influence
the rank, as indicators they are noisy and force us to collect
a huge number of samples.
To maximize the effect on the rank of our embeddings, we need to choose
$S$ with a very specific rank distribution, e.g. with matrices of
extremely low rank (while keeping all rows distinct).\\
A report on some experimental results can be found in \cite{CGC-cry-prep-BRS10}.


\section{Further remarks and other results}
The first subsection contains some results on how our representation could achieve a weaker notion of linearity.

In Subsection \ref{OtherRepresentations} we report other thinkable representations, that unfortunately are impractical.
The main objective in these constructions is to identify the right compromise between computational feasibility and quantity of information that can be obtained.

Then, in Subsection \ref{Further remark} we prove the fact, using classical and easy arguments, that 
it is unlikely to embed the AES cipher into a linear cipher, unless one uses a huge-dimensional vector space (and so this embedding is useless in practice).

\subsection{On a weaker notion of linearity}
\label{Weakernotion}
The results in this section are jointly with L. Maines and the proofs are contained in her Master's thesis \cite{CGC-tesi5-maines} (see also \cite{CGC-alg-art-maines}), supervised by the second author.\\
The main goal sought in Section \ref{preliminary rep}, Section \ref{FirstRepresent}, Section \ref{OrbitRepresentation}, and Section \ref{OtherRepresentations} 
is to find practical embedding of $(\FF_{2})^{128}$ into a larger space where all components of the round function become linear. 
This is impossible, as shown in Section \ref{Further remark}, but what we achieve in Section \ref{OrbitRepresentation} is an embedding where the non-linear maps are ``not so far'' from linear maps. 
There are many notions of ``non-linearity'', but none of them can be easily computed in our setting. When we say ``not so far from linear'', 
we mean that these functions behave with matrix ranks in a way similar to that of linear maps, as discussed in Section \ref{strategies}. 
However, we have been able to introduce a new non -linearity notion, that we call {\em $s$-extendibility} (Definition \ref{s-extendible}). We are not able to apply it in the embedding
\begin{equation}
\label{right}
\alpha:v\rightarrow(\varepsilon(v),\varepsilon ({\bf M} v), \cdots, \varepsilon ({\bf M}^7 v)).
\end{equation}
but we can apply it\footnote{under specific conditions on $M$} to
\begin{eqnarray*}
\alpha:v\rightarrow(\varepsilon(v),\varepsilon ({\bf M} v)).
\end{eqnarray*}
and so our definition and our results (the main results of this section is Theorem \ref{th:generalizzazione}) should be seen as a step forward the complete understanding of the surviving non-linearity in (\ref{right}).

\begin{definition}
\label{s-extendible}
Let $V=(\FF_2)^{r}$ and  $W=(\FF_2)^{s}$, with $s > r$.
Let $\sigma \in \Sym(V)$ and $\alpha$ be an injective map $\alpha: V \rightarrow W$.
We say that $\sigma$ is {\bf $s$-extendible} (via $\alpha$) if $\forall \{v^h\}_{1\leq h\leq s} \subset V$ we have 
$$\sum_{h=1}^s {\alpha(v^h)}=0 \iff \sum_{h=1}^s{\alpha(\sigma(v^h))}=0.$$ 
\end{definition}

\begin{remark}
\label{doppioni}
If $v^1=v^2$ and $v^3=v^4$, then $\forall \alpha$ and $\forall \sigma$ we have
$$
\alpha(v^i)+\alpha(v^2)+\alpha(v^3)+\alpha(v^4)=0
$$
and 
$$
\alpha(\sigma(v^i))+\alpha(\sigma(v^2))+\alpha(\sigma(v^3))+\alpha(\sigma(v^4))=0.
$$
So if we test the $4$-extendibility of $\sigma$ only on these sets of vectors, we will find that any $\sigma$ is $4$-extendible. We call these vectors 
``coupled vectors''. 
\end{remark}

We note that if  $\sigma$ is  $s$-extendible  $\forall s \in \N$, then  $\sigma$ is  linearly  extendible, according to Definition \ref{linearly-extendible}. Moreover, any linear map is $s$-extendible for all $s$. A random map is a $2$-extendible but (with high probability) it is
not $s$-extendible for any $s\geq 4$. Therefore, any $4$-extendible map can be considered closer to a linear map. We would like to have results on our embedding concerning the $s$-extendibility of maps.
A first result in this direction is obtained using the space embedding 
\begin{equation}
\label{partialorbita}
\alpha(v)=(\varepsilon(v),\varepsilon ({\bf M} v)),
\end{equation}
where ${\bf M}$ is a $(n \times n)$-matrix with entries in $\FF_{2^m}$, as we are going to explain. 

\begin{definition}\label{def:amicigen4}
 Let $i,j,x,y,\alpha,\beta,\dots \in \FF_{2^m}$ and $\mathbf{M}$ an $(n \times n)$-matrix with entries in $\FF_{2^m}$ 
$$
{\bf M}=\left(\begin{array}{cccc}
m_{11} & m_{12} & \ldots & m_{1n}\\
m_{21} & m_{22} & \ldots & \ldots \\
\vdots & \vdots & \ddots & \ddots \\
m_{n1} & \ldots & \ldots & m_{nn}
\end{array}\right).
$$

The vectors $w_1,w_2,w_3,w_4 \in (\FF_{2^m})^{2n}$ are $4$-{\sf related vectors} if they can be permuted in order to have the following form:
$$\begin{array}{cc|cccccc}
 & & 1 & 2 & \dots & n+1  & \dots& 2n \\
\cline{2-8}
\mathbf{1} & w_1, & (i, & x, & \dots & m_{11}i+m_{12}x+\dots,  & \dots & m_{n1}i+m_{n2}x+\dots) \\
\mathbf{2} & w_2, & (i, & y, & \dots & m_{11}i+m_{12}y+\dots,  & \dots & m_{n1}i+m_{n2}y+\dots) \\
\mathbf{3} & w_3, & (j, & x, & \dots & m_{11}j+m_{12}x+\dots,  & \dots & m_{n1}j+m_{n2}x+\dots) \\
\mathbf{4} & w_4, & (j, & y, & \dots & m_{11}j+m_{12}y+\dots,  & \dots & m_{n1}j+m_{n2}y+\dots) \\
\end{array}
$$
\end{definition}

\ \\
Four related vectors $w_1,\ldots,w_4$ are admissible vectors
$ \alpha(v_1)=(\varepsilon(v_1),\varepsilon(\mathbf{M} v_1))$,
 $\alpha(v_2)=(\varepsilon(v_2),\varepsilon(\mathbf{M} v_2))$,
$\alpha(v_3)=(\varepsilon(v_3),\varepsilon(\mathbf{M} v_3))$,
$\alpha(v_4)=(\varepsilon(v_4),\varepsilon(\mathbf{M} v_4))$
such that
$$\varepsilon(v_1)+\varepsilon(v_2)+\varepsilon(v_3)+\varepsilon(v_4)=0\, ,$$
but we do not know the sum 
$\varepsilon(\mathbf{M} v_1)+\varepsilon(\mathbf{M} v_2)+\varepsilon(\mathbf{M} v_3)+\varepsilon(\mathbf{M} v_4)$.

Let $\sigma$ be a parallel maps over $(\FF_{2^m})^{2n}$. The image of $4$-related vectors via $\sigma$ can be seen as
\pagebreak
{\footnotesize $$\begin{array}{cc|ccccccc}
 & & 1 & 2 & \dots & n+1 & \dots& 2n \\
\cline{2-8}
\mathbf{1} & w_1,   & (\sigma(i), & \sigma(x), & \dots & m_{11}\sigma(i)+m_{12}\sigma(x)+\dots,  & \dots & m_{n1}\sigma(i)+m_{n2}\sigma(x)+\dots) \\
\mathbf{2} & w_2^*, & (\sigma(i), & \sigma(y), & \dots & m_{11}\sigma(i)+m_{12}\sigma(y)+\dots,  & \dots & m_{n1}\sigma(i)+m_{n2}\sigma(y)+\dots) \\
\mathbf{3} & w_3^*, & (\sigma(j), & \sigma(x), & \dots & m_{11}\sigma(j)+m_{12}\sigma(x)+\dots,  & \dots & m_{n1}\sigma(j)+m_{n2}\sigma(x)+\dots) \\
\mathbf{4} & w_4^*, & (\sigma(j), & \sigma(y), & \dots & m_{11}\sigma(j)+m_{12}\sigma(y)+\dots,  & \dots & m_{n1}\sigma(j)+m_{n2}\sigma(y)+\dots) \\
\end{array}
$$}

\begin{definition}\label{def:doppioni4}
 $4$-related vectors $w_1,\ldots,w_4$ are {\sf totally} $4$-related if $$w_1+w_2+w_3+w_4=0.$$
\end{definition}
\begin{definition}
Given $(x,y,z,a,b,c) \in \NN ^6$ and an $(n\times n)$-matrix $\mathbf{M}$, we say that  $(x,y,z,a,b,c)$ {\sf fits} $\mathbf{M}$ 
if the following sums of elements of $\det(\mathbf{M})$ are non-zero:  
\begin{itemize}
 \item the sums having a number of elements equal to 
\begin{table}[!h]\resizebox{\columnwidth}{!}{
\begin{tabular}{ccc}
 $\displaystyle\sum_{i=0}^x \binom{n-c}{i}\binom{n-b}{x-i}\binom{b-i}{y}x!y!$ & 
 $\displaystyle\sum_{i=0}^x \binom{n-b}{i}\binom{n-c}{x-i}\binom{c-i}{z}x!z!$ &
 $\displaystyle\sum_{i=0}^y \binom{n-a}{i}\binom{n-c}{x-i}\binom{c-i}{z}y!z!$ \\
 \textit{when} $z=0,x\neq 0,y\neq 0$ & \textit{when} $y=0,x\neq 0,z\neq 0$ & \textit{when} $x=0,y\neq 0,z\neq 0$
\end{tabular}}
\end{table}
 \item the sums having a number of elements equal to
 $$\small \sum_{i=0}^{x}\sum_{j=0}^{y}\binom{n-c}{i}\binom{n-b}{x-i}\binom{n-a}{j}\binom{(n-c)-i}{y-j}\binom{c-(x-i)-j}{z}x!y!z!$$ 
when $x\neq0,y\neq0,z\neq0$.
\end{itemize}
\end{definition}
The main result of this section is the next theorem that gives sufficient conditions on ${\bf M}$ in order to make all $\sigma:V \rightarrow V$ into $4$-exendible maps.
\begin{theorem}\label{th:generalizzazione}
Let $\mathbf{M}$ be an $(n\times n)$-matrix, with entries in $\FF_{2^m}$ such that:
\begin{enumerate}
 \item $\det(\mathbf{M})\neq 0$;
 \item all the $k\times k$ minors are non-zero  ($0<k<n$);
 \item all sextuple $(x,y,z,a,b,c)$ such that
\begin{itemize}
 \item $0< a,\ b,\ c \leq n$;
 \item $a+b+c=2n$;
 \item $a\geq b\geq c$;
 \item $0\leq x,y,z \leq n$;
 \item $x+y+z=n$;
 \item $x<a$, $y<b$, $z<c$;
\end{itemize}
  fit $\mathbf{M}$.
\end{enumerate}
Then any $4$-related vectors are  totally related if and only if they are coupled.
\end{theorem}

Thanks to Theorem \ref{th:generalizzazione} and Remark \ref{doppioni}, we have the following 
\begin{corollary}
In the hypothesis of Theorem \ref{th:generalizzazione}, any map is $4$-extendible.
\end{corollary}

\subsection{Other embeddings of this kind}
\label{OtherRepresentations}
We can also build other embeddings similar to those described in previous sections. 
The main objective in these constructions is to identify the right compromise between computational feasibility and quantity of information that can be obtained.
In Section \ref{FirstRepresent}, we constructed the embedding $\varepsilon$ that has been useful to make linear the $S$-box maps which are the classical non-linear maps of a cryptosystem. We had to abandon the linearity of {\sf MixColumns} (for AES) and the {\sf pLayer} (in case of PRESENT).
In order to use some more information about the {\sf MixColumns} (or the pLayer for PRESENT), we have considered the embedding given in Section \ref{OrbitRepresentation}:
$$\alpha(v)=(\varepsilon(v),\varepsilon ({\bf M} v),\dots, \varepsilon({\bf M}^{t-1} v)),$$
where ${\bf M}$ is the full Mixing Layer.
The strength of this enbedding is that we can exploit the low order of {\bf M} to force the linearity of {\bf M}.
The disadvantages are that we have lost some computational efficiency and that the $S$-box is non-linear again (but with a lower non-linearity).

For AES, we considered also the embedding given by 
$$\alpha(v)=(\varepsilon(v),\varepsilon ({\sf MC}(v)),\dots, \varepsilon({\sf MC}^{3} (v))),$$ 
since the order of the {\sf MixColumns} is equal to $4$ and the {\sf MixColumns} operation was the only to be non-linear in Section \ref{FirstRepresent}. 
Unfortunately, in this context both the {\sf ShiftRows} and the parallel maps are non-linear and so we put aside this idea.

Although the following two embeddings could provide a lot of information about a cryptosystem, 
\begin{itemize}
\item  $\alpha(v)=(\varepsilon(v),\varepsilon (({\bf M}\circ {\rm Sbox})v),\dots, \varepsilon(({\bf M}\circ {\rm Sbox})^{t-1} v)) 
\quad (t={\sf o}({\bf M}\circ {\rm Sbox}) )$
\item  $\alpha(v)=(\varepsilon(v),\varepsilon ((\gamma \lambda \sigma_k)v),\dots, \varepsilon((\gamma \lambda \sigma_k))^{t-1} v)) 
\quad (t={\sf o}(\gamma \lambda \sigma_k))$
\end{itemize}
they are very impractical, since the order of $({\bf M}\circ {\rm Sbox})$ and of $(\gamma \lambda \sigma_k)$ is huge.

\subsection{On complete linearizations of AES}
\label{Further remark}
Let ${\mathcal C}$ be any block cipher such that the plain-text space $\MM$
coincides with the cipher space. Let ${\mathcal K}$ be the key space.
Any key $k\in {\mathcal K}$ induces a permutation $\tau_k$ on $\MM$.
Since $\MM$ is usually $V=(\FF_2)^n$ for some $n\in \N$, we can consider
$\tau_k\in \Sym(V)$.
We denote by $\Gamma=\Gamma({\mathcal C})$ the subgroup of $\Sym(V)$ generated
by all the $\tau_k$'s.
Unfortunately, the knowledge of $\Gamma({\mathcal C})$ is out of reach
for the most important block ciphers, such as 
the AES \cite{CGC-misc-nistAESfips197} and the 
DES \cite{CGC-misc-nistDESfips46}. 
However, researchers have been able to compute another related
group. Suppose that ${\mathcal C}$ is the composition of $l$ rounds
(the division into rounds is provided in the document describing the cipher).
Then any key $k$ would induce $l$ permutations, 
$\tau_{k,1},\ldots,\tau_{k,l}$, whose composition is $\tau_k$.
For any round $h$, we can consider $\Gamma_h({\mathcal C})$ as the 
subgroup of $\Sym(V)$ generated by the $\tau_{k,h}$'s 
(with $k$ varying in ${\mathcal K}$).
We can thus define the group $\Gamma_{\infty}=\Gamma_{\infty}({\mathcal C})$
as the subgroup of $\Sym(V)$ generated by all the $\Gamma_h$'s.
Obviously, $\Gamma \,\leq\, \Gamma_{\infty} \,.$
Group $\Gamma_{\infty}$ is traditionally called the \emph{group generated
by the round functions} with  independent sub-keys.
This group is known for some important ciphers, for example we have

\begin{proposition}[\cite{CGC-cry-art-sparr08},\cite{CGC-cry-art-Wernsdorf1}]
\label{wer}

 $$\Gamma_{\infty}({\mathrm{AES}})=\Alt((\FF_2)^{128}).$$
\end{proposition}

It is very likely (and it is common belief among researchers) that $\Gamma_{AES}=\Gamma_{\infty}({\mathrm{AES}})=\Alt((\FF_2)^{128})$.
Assuming this, we discuss in this section the possibility of viewing
 $\Gamma_{AES}$ as a subgroup of $\mathrm{GL}(V)$ with $V$ of small dimension.
In Cryptography it is customary to present estimates as powers of two, so our problem becomes to find the smallest $\ell$ such that $\Gamma_{AES}$
can be linearized in $\GL((\FF_2)^{2^{\ell}})$.
A classical proof is given in \cite{CGC-alg-art-Wagner76} that $\ell=128$.
We feel desirable to obtain a result with a simpler proof.
Our estimate is weaker than Wagner's, but strong enough to show the linearization infeasibility.

There are two obvious ways to show that a finite group $A$ cannot be contained (as isomorphic image) in a finite group $B$. The first is to show that $|A|>|B|$, the second is to show that there is $\eta \in A$ such that its order is strictly larger than the maximum element order in $B$.
Subsection \ref{first approach} presents our result using the first approach and we show that $\ell \geq 67$, which is more than enough to ensure the infeasibility of the linearization attack. This subsection's argument is completely elementary. Subsection \ref{second approach} present our result using the second approach and we show again that $\ell \geq 67$. It is interesting that, although here some more advanced argument is needed (results in number theory), we reach the same estimate.

\subsubsection{First approach}
\label{first approach}
In this subsection we show that the order of $\Alt((\FF_2)^{128})$ is strictly larger than the order of $\GL(V)$, with $V=(\FF_2)^{2^{66}}$, so that $\ell \geq {67}$.\\
\\  
We begin with showing a lemma.

\begin{lemma}
\label{lemma-final}
The following inequality holds
$$ 2^{(2^7)^{19}}< 2^{128}!< 2^{(2^7)^{20}}.$$
\begin{proof}

Let $n=2^7$, we have to show $ 2^{n^{19}}< 2^n!< 2^{n^{20}}.$ 
We first show that $ 2^{n^{19}}< 2^n!$.
The following inequality holds for $1\leq i \leq n-2$ and $ 1 \leq h \leq 2^{n-i}$ 
\begin{equation}
\label{eqw}
\frac{1}{2^{n-i}}\geq \frac{1}{2^{n-i+1}-h} \;.
\end{equation}
Clearly
\begin{eqnarray*}
2^{n}! >2^{n^{19}} & \iff & 2^{n}(2^{n}-1)!> 2^{n} \cdot 2^{n^{19}-n}\\
                   & \iff & (2^{n}-1)(2^n-2)! > 2^{n^{19}-n}\cdot \frac{2^{n}-1}{2^{n}-1} \quad .
\end{eqnarray*}
We apply (\ref{eqw}) with $i=1$ and $h=1$ and so we must prove
$$(2^{n}-1)(2^n-2)! >2^{n^{19}-n}\cdot \frac{2^{n}-1}{2^{n-1}},$$
i.e.  $(2^n-2)! >  2^{n^{19}-n-(n-1)}$.
We use the same inequality for all $2 \leq h \leq 2^{n-1}$ and 
we obtain that we must verify
$(2^{n-1}-1)! >2^{n^{19}-n-2^{n-1}(n-1)}$.
Then we proceed by applying (\ref{eqw}) for all $2 \leq i \leq n-2$ 
and all $ 1 \leq h \leq 2^{n-i}$, so that we need only to prove
 $$
  (2^{n-(n-1)}-1)!\geq  2^{n^{19}-n-\sum_{i=1}^{n-1}2^{n-i}(n-i)} \,.
 $$
In other words, we have to prove  
\begin{equation}
\label{right-size}
1 > 2^{n^{19}-n-\sum_{i=1}^{n-1}2^{n-i}(n-i)}, \quad \textrm{that is,} \quad 0> n^{19}-n-\sum_{i=1}^{n-1}2^{n-i}(n-i).
\end{equation}
But a direct check shows that the right-hand size of (\ref{right-size}) 
holds when $n = 2^{7}$.\\ 

We are left to demonstrate the following inequality: $2^n!< 2^{n^{20}}$.\\
We proceed by induction for $2\leq n\leq 2^7$. In this range a computer computation shows that 
  \begin{equation}
  \label{b}
  n^{20} + 2^n n +2^n < (n+1)^{20}.
  \end{equation}
  
When $n=2$, we have $2^2!< 2^{2^{20}}$.
Suppose that $2^n!< 2^{n^{20}}$ and $n\leq 2^7$. 
We have to prove that $2^{(n+1)}!< 2^{(n+1)^{20}}$.
Since $ 2^{n+1}! \,=\,  (2^n \cdot 2)!  = 2^n!(2^n+1) \cdots (2^n+2^n)$, 
we have
 $$
 2^n!(2^n+1) \cdots (2^n+2^n)  < 2^{n^{20}+n+1}\cdot (2^n +2) \cdots (2^n+2^n) 
         \leq \, 2^{n^{20}+2^n(n+1)} = 2^{n^{20}+2^n n +2^n}
 $$
 and, applying ($\ref{b}$), we get $2^{n^{20}+2^n n +2^n}<2^{(n+1)^{20}}.$\\
 Then the claimed inequality $2^{n+1}!< 2^{(n+1)^{20}}$ follows.
 \end{proof}
\end{lemma}

Our result is contained in the following proposition.
\begin{proposition}
\label{dimensionTOT}
Let $V=(\FF_2)^{2^{\ell}}$ with $\ell \geq 2$. If $G < \mathrm{GL}(V)$, with $G$ isomorphic to $\Alt((\FF_2)^{128})$,  then $\ell \geq 67$.
\begin{proof}
If $G<\GL(V)$, then $|G|\leq |\GL(V)|$. But $|\Sym((\FF_2)^{128})|=2^{128}!>2^{2^{133}}$ thanks to Lemma \ref{lemma-final} and so 
$$
  |G|=|\Alt((\FF_2)^{128})|= \frac{|\Sym((\FF_2)^{128})|}{2} > \frac{2^{2^{133}}}{2}
  = 2^{2^{133}-1}> 2^{2^{132}} > |\GL((\FF_2)^{2^{66}})| \,.
$$
Therefore, $\ell =66$ is not large enough.
\end{proof}
\end{proposition}
\begin{remark}
\label{werns}
We could improve the previous bound to $\ell\geq 68$ by using the finite version of the Stirling
fomula:
$$
   n {\mathrm{log}}_2 n  -n{\mathrm{log}}_2(e) \leq {\mathrm{log}}_2 (n!) 
   \leq n {\mathrm{log}}_2 n  -n{\mathrm{log}}_2(e) + {\mathrm{log}}_2 n ,
   \quad {\left(\frac{n}{e}\right)}^n \leq n! \leq n \left({\frac{n}{e}}\right)^n \,.
$$
\end{remark}

\subsubsection{Using the order of the elements}
\label{second approach}
In this subsection we compare the maximum order of elements in the two groups $\Alt((\FF_2)^{128})$ and $\GL((\FF_2)^{2^{\ell}})$. We use permutations of even order. We denote by ${\sf o}(\sigma)$ the order of any permutation $\sigma$, with $\sigma \in \Alt((\FF_2)^{128})$ or $\sigma \in \GL((\FF_2)^{2^{\ell}})$.

The best available result for $\GL((\FF_2)^{2^{\ell}})$ is given by the following theorem

\begin{theorem}[\cite{CGC-alg-art-Darafsheh08}]
\label{GLelements}
Let $\sigma \in \GL((\FF_2)^N)$, with  ${\sf o}(\sigma)$ is even and $N\geq 4$. Then 
$${\sf o}(\sigma) \leq 2(2^{N-2}-1)=2^{N-1}-2.$$

Moreover, there is $\sigma \in \GL((\FF_2)^N)$ whose order attains the upper bound.
\begin{proof}
It comes directly from Theorem 1 in \cite{CGC-alg-art-Darafsheh08}, with $p=q=2$ and $N \geq 4$ (so point (a) and (b) do not apply).
\end{proof}
\end{theorem}

As regards the order of the elements in $\Alt((\FF_2)^{128})$, we would like to use the following theorem 

\begin{theorem}[\cite{CGC-alg-book-DixonMortimer96}]
\label{Altelement}
Let $\nu\geq 3$ and $n=2^\nu$. Then $\Alt((\FF_2)^{\nu})$ contains an element $\eta$ of 
order (strictly) greater then $e^{\sqrt{(1/4)n \ln n}}$.
\end{theorem}

The previous theorem is the special case of Theorem 5.1.A at p.145 
in  \cite{CGC-alg-book-DixonMortimer96} when $q=2$.

In order to be able to compare the two estimates coming from Theorem \ref{GLelements} and Theorem \ref{Altelement}, we rewrite Theorem \ref{Altelement} as follows, in order to have ${\sf o}(\eta)$ even. Our proof is an easy adaption of the proof contained in \cite{CGC-alg-book-DixonMortimer96}.
\begin{theorem}
\label{evenAltelement}
Let $\nu\geq 7$ and $n=2^\nu$. Then $\Alt((\FF_2)^{\nu})$ contains an element $\eta$ 
with ${\sf o}(\eta) > e^{\sqrt{(1/4)n \ln n}}$ and ${\sf o}(\eta)$ even.
\begin{proof}
Let $z$ be a prime number such that
$ 4+\sum_{3 \leq p\leq z}p \leq n\,$,
where the sum runs over (distinct odd) prime numbers.
Then $\Alt((\FF_2)^{\nu})$ contains an element $\eta_z=\sigma \sigma' \sigma_3 \cdots \sigma_p \cdots \sigma_z$ such that: $\sigma$ and $\sigma'$ are transpositions,
$\sigma_p$ is a cycle of length $p$, and all cycles $\{\sigma,\sigma',\sigma_3,\ldots,\sigma_z\}$ act on disjoint subsets of $(\FF_2)^{\nu}$.
In other words, the non-trivial cycles of $\eta_z$ are two transpositions and some  cycles with length $3,\ldots, z$.
As a consequence, the order of $\eta_z$ is  $2 \prod_{3 \leq p \leq z} p$.

We are going to show that there is $z \in \NN$ such that
$$
4+\sum_{2<p\leq z}p \leq n \quad \quad \textrm{and} \quad \quad (\vartheta(z))^2 > \frac{1}{4}n\ln(n) 
$$
where $\vartheta(z)=\ln({\sf o}(\eta_z))=\ln(2)+\sum_{2<p\leq z}\ln(p)$; in the following we consider $\vartheta^{*}(z)=\vartheta(z)-\ln(2)=\sum_{2<p\leq z}\ln(p)$.


Since $n \geq 2^7$,  we note that
$4+\sum_{2<p\leq 19} p = 79 < 128\leq n.$

Let $f(z)=\frac{z}{\ln(z)}$. Since $f(z)$ is an increasing function for real $z >e$, in case $z$ is real and $z\geq 19,$ we have that

 \begin{equation} \label{B}
 f(4)\ln(4)+f(3)\ln(3)=7<f(19)\ln(3)\leq f(z) \ln(3)  
\end{equation}
and so we can write (if  $z \geq 19$ and $z \in \RR$)
\begin{eqnarray*}
4+ \sum_{2<p\leq z}p&=&f(4)\ln(4)+\sum_{2<p\leq z}f(p)\ln(p)\\
                    &=&f(4)\ln(4)+f(3)\ln(3)+\sum_{3<p\leq z}f(p)\ln(p)\\
                    &< &f(z)\ln(3)+\sum_{3<p\leq z}f(z)\ln(p)\\
                    &=& \sum_{2<p\leq z}f(z)\ln(p) = f(z)\sum_{2<p\leq z}\ln(p)=f(z)\vartheta^{*}(z).
\end{eqnarray*}



We shall choose  $\bar z\geq 19$ such that $f(\bar z)\vartheta^{*}(\bar z)=n$. Such a $\bar z$ exists because $f(19)\vartheta^{*}(19)<100<n$ and 
 $f(z)\vartheta^{*}(z)$ is an increasing function assuming all values.\\
Since $\vartheta^{*}(z)>z/2$ for all $z\geq 19$, we have
$$
  n=\frac{\bar z\vartheta^{*}(\bar z)}{\ln(\bar z)}<\frac{2(\vartheta^{*}(\bar z))^2}{\ln(2\vartheta^{*}(\bar z))}=\frac{4(\vartheta^{*}(\bar z))^2}{2\ln(2\vartheta^{*}(\bar z))}=
  f(4(\vartheta^{*}(\bar z))^2)\,.
$$
However we also have $f(n \ln(n))<n$.
Since $f$ is an increasing function, this shows that $n \ln(n)<4(\vartheta^{*}(\bar z))^2 < 4(\vartheta(\bar z))^2$.
It is now enough to consider $\tilde z$ as the largest prime smaller than $\bar z$.
\end{proof}
\end{theorem}   
Now, we compare the estimate from  Theorem \ref{GLelements} and Theorem \ref{Altelement}.\\
Take $n=2^{128}$ and $\eta \in \Alt((\FF_2)^{128})$ such that ${\sf o}(\eta)\geq e^t$ \, (${\sf o}(\eta)$ even), 
where $t=\sqrt{(1/4)n \ln n}=\sqrt{(1/4)2^{128} \ln (2^{128})}$.\\
Since
$$
e^t=e^{\sqrt{2^{126} 128 \ln 2}}=e^{\sqrt{2^{133} \ln 2}}=(e^{\sqrt{2 \ln 2}})^{2^{66}},
$$
by replacing $e$ with $2^{\log_2 e}$, we obtain
$$
e^t=(2^{(\log_2 e) \sqrt{2 \ln 2}})^{2^{66}}=2^{2^{66} (\log_2 e) \sqrt{2 \ln 2}}=2^{2^{66}\varepsilon},
$$
where $\varepsilon \in \RR$ is circa $1.69$. According to Theorem \ref{evenAltelement}, 
the order of $\eta$ is at least  ${\sf o}(\eta)\geq e^{2^{66}\varepsilon}$.
If $\Alt((\FF_2)^{128})\subset \GL((\FF_2)^N)$, we then need the the smallest $N$ such that 
${\sf o}(\eta) \leq (2^{N-1}-2)$ (Theorem \ref{GLelements}).  
In other words we have to see when the following inequality holds
\begin{equation}
\label{order bound}
{\sf o}(\eta)=2^{2^{66}\varepsilon} \leq 2^{N-1}-2.
\end{equation}

We observe that
\begin{itemize}
\item if $N=2^{66}$, then (\ref{order bound}) is false, since $2^{2^{66}\varepsilon}> 2^{2^{66}}> 2^{2^{66}-1}-2$;
\item if $N=2^{67}$, then (\ref{order bound}) is true, since $2^{2^{66}\varepsilon}< 2^{2^{66}(1.7)}< 2^{2^{67}-1}-2$.
\end{itemize}

Therefore, we need at least $\ell \geq 67$ in order to embed $\Alt((\FF_2)^{128})\subset \GL(V)$, which is exactly the same value as in 
Proposition \ref{dimensionTOT}.

\begin{remark}
It is shown in Landau \cite{CGC-alg-art-landau1903} that the maximum order of an element in $\Sym((\FF_2)^{\nu})$ is asymptotic to $e^{\sqrt{n \ln n}}$ as $n \rightarrow \infty$ (with $n=2^\nu$).
Assuming this, we observe that we could slightly improve the value of $\ell$ we need to $\ell \geq 68$,
which is the same as Remark \ref{werns}.
\end{remark}


\section*{Acknowledgments}
A large part of these results comes from the first author's Ph.D thesis, after some initial insights by the third author.
The first author would like to thank the second author (her supervisor).

These results have been presented in a few talks (2007: Trento; 2008: Cork, Pisa; 2009: Trento; 2010: Marseille, Torino) and several scientific discussions with colleagues. 
The authors would like to thank the following people for their valuable comments and suggestions: 
G. Bertoni,  A. Caranti, F. Dalla Volta, O.~Dunkelman, P. Fitzpatrick, P. Fragneto, P. Gianni, L. Maines, T. Mora, L.~Perret, C. Traverso, R. Wernsdorf.

For their help in the attack implementation the authors thank E. Bertolazzi and F. Caruso.

The initial discussion about this work has been supported by the STMicroelectronics contract `` Complexity issues in algebraic Coding Theory and Cryptography''. Further discussion took place during the Special Semester on Groebner Bases (2006), organized by RICAM, {\em Austrian Academy of Sciences} and RISC, Linz, Austria.

Part of this research has been funded by: {\tt Provincia Autonoma di Trento grant}{\em ``PAT-CRS grant''}, {\tt MIUR grant}{\em ``Algebra Commutativa, Combinatoria e Computazionale''}, 
{\tt MIUR grant} {\em  ``Rientro dei Cervelli''}.

\bibliography{RefsCGC}

\end{document}